\documentclass[11pt,a4paper]{article}
\usepackage[a4paper,left=2cm,right=2cm,top=2.5cm,bottom=3.5cm]{geometry}
\usepackage[utf8]{inputenc} 
\usepackage[T1]{fontenc}    
\usepackage{hyperref}       
\usepackage{booktabs}       
\usepackage{amsfonts}       
\usepackage{lipsum}
\usepackage{lipsum}
\usepackage{graphicx}
\usepackage{xcolor}
\usepackage{framed}
\usepackage{bm}
\usepackage{tcolorbox}
\usepackage{amssymb}
\usepackage{amsmath}
\usepackage{url}

\usepackage[]{eurosym}
\usepackage[braket, qm]{qcircuit}
\usepackage{framed}
\usepackage{subfig}
\usepackage{caption}
\usepackage{titlesec}
\usepackage{authblk}
\usepackage[linesnumbered,ruled,vlined]{algorithm2e}
\RestyleAlgo{boxruled}
\usepackage[
    natbib=true,
    style=nature,
    sorting=none,
    giveninits=false
]{biblatex}
\newcommand{\hrulealg}[0]{\vspace{1mm} \hrule \vspace{1mm}}

\title{Qualifying quantum approaches for hard industrial optimization problems. A case study in the field of smart-charging of electric vehicles}
\author[1,2]{Constantin~Dalyac}
\author[1]{Loïc~Henriet\thanks{corresponding author : loic@pasqal.io}}

\author[3]{Emmanuel~Jeandel}
\author[4,5]{Wolfgang~Lechner}
\author[3]{Simon~Perdrix}
\author[6]{Marc~Porcheron}
\author[3,6]{Margarita~Veshchezerova}
\affil[1]{\small Pasqal, 2 avenue Augustin Fresnel, 91120 Palaiseau, France}
\affil[2]{\small LIP6, CNRS, Sorbonne Université, 4 Place Jussieu, 75005 Paris, France}

\affil[3]{\small LORIA, CNRS, Université de Lorraine, Inria Mocqua, F 54000 Nancy, France}
\affil[4]{\small University of Innsbruck, Technikerstrasse 21a, Innsbruck, Austria}
\affil[5]{\small Parity Quantum Computing GmbH, Rennweg 1, Innsbruck, Austria}
\affil[6]{\small EDF R\&D, 7, boulevard Gaspard Monge, 91120 Palaiseau, France}

\begin{document}

\maketitle
\begin{abstract}
In order to qualify quantum algorithms for industrial NP-Hard problems, comparing them to available polynomial approximate classical algorithms and not only to exact ones -- exponential by nature -- , is necessary. This is a great challenge as, in many cases, bounds on the reachable approximation ratios exist according to some highly-trusted conjectures of Complexity Theory.  An interesting setup for such qualification is thus to focus on particular instances of these problems known to be "less difficult" than the worst-case ones and for which the above bounds can be outperformed: quantum algorithms should perform at least as well as the conventional approximate ones on these instances, up to very large sizes.
We present a case study of such a protocol for two industrial problems drawn from the strongly developing field of smart-charging of electric vehicles. Tailored implementations of the Quantum Approximate Optimization Algorithm (QAOA) have been developed for both problems, and tested numerically with classical resources either by emulation of Pasqal's Rydberg atom based quantum device or using Atos Quantum Learning Machine. In both cases, quantum algorithms exhibit the same approximation ratios than conventional approximation algorithms, or improve them. These are very encouraging results, although still for instances of limited size as allowed by studies on classical computing resources. The next step will be to confirm them on larger instances, on actual devices, and for more complex versions of the problems addressed.
\end{abstract}
\section{Introduction}

In the quest for a quantum advantage, situating quantum algorithms and technologies with respect to classical approaches for solving NP-hard problems is a key question. Seminal works from Shor and Grover were a major breakthrough in this perspective in the early developments of quantum computing. Based on the theoretical model of quantum computation, they developed quantum algorithms that provide respectively an exponential speed-up for integer factorization \,\cite{Shor94}, and a quadratic speed-up on exhaustive search in an unstructured database \,\cite{Grover96}, a principle applicable to any NP problem providing  at most such a speed-up in terms of query complexity\,\cite{Bennett_1997,viamontes2005quantum}. 
However, implementing these algorithms requires fault-tolerant quantum machines handling a large number of qubits, which have yet to be built. \\

While many technological obstacles currently impede the creation of such machines, experimental physicists have been capable of controlling quantum systems precisely enough to simulate complex many-body quantum systems. These quantum devices present strong quantum properties and offer scientists a control on the quantum aspects of physical systems. They can have sizes of several hundreds of quantum particles, and because of the unavoidable coupling between the system and its environment, these quantum platforms fall in the category of Noisy Intermediate Scale Quantum (NISQ) devices\,\cite{Preskill_2018}. Among them, there is a strong belief that Analog Quantum Simulators (AQS) can perform specific tasks intractable for classical computers in polynomial time, such as the dynamical simulation of strongly interacting quantum Hamiltonians\,\cite{Choi_2016,Scholl20}, and it is expected that AQS will be among the first to propose useful applications in the short-term\,\cite{Kendon2020}. Lately, there has been growing interest in knowing if the quantum characteristics of these devices can be steered towards outperforming classical computers on industry-relevant tasks. An active field of research is currently guided towards combinatorial optimization, where the Hilbert space spanned by the many-body quantum system is used to efficiently encode a high-dimensional discrete problem. \\

In this context, algorithms that can run on such NISQ devices have been developed. Quantum Annealing \,\cite{Morita08}, the Quantum Adiabatic Algorithm \,\cite{Farhi00}, and the Quantum Approximate Optimization Algorithm \,\cite{Farhi14} are among the most promising ones. A better understanding of the performances of these approaches on industrial NP-hard problems is of great interest, both for quantum computing adoption and for the application domains concerned.\\

Approximate results are of great interest for practical applications, specifically industrial ones where a close-to-optimal solution is of significant value when exact solutions are unreachable due to exponential conventional computing time. Interestingly, some NP-hard problems are easier to approximate than others. For example, while the $0/1$-Knapsack problem is NP-complete, it admits a fully polynomial-time approximation scheme (PTAS) \,\cite{Ibarra75}. To grasp the performances and the quality of solutions provided by NISQ algorithms, we must rigorously compare them to their approximate classical counterparts. Determining where quantum approaches sit in the approximation complexity landscape is key in understanding their potential for practical problems. One should be aware that for some NP-hard problems bounds have been proven on these ratios which cannot be exceeded by classical algorithms unless some very strongly-believed conjectures of Complexity Theory would be broken; thus, improving these ratios beyond these bounds in these cases would establish a quantum supremacy for solving some NP-hard problems. Another desirable scenario is that quantum approaches find comparable approximation ratios as classical algorithms but faster or at a lower cost, which would already provide significant value for industries. \\

Taking this latter point into consideration, an interesting workout for qualification of quantum approaches is to focus on particular instances of NP-Hard problems known to be "less-difficult" than the worst-case ones, in the sense that approximation algorithms do exist for them that could outperform the above bounds, which are precisely established for the worst-case instances : our quantum algorithms should perform at least as well as the conventional approximate ones on these instances, up to very large sizes. \\

In this article, we present and discuss a case study based on this protocol, for two problems drawn from the rapidly growing sector of smart-charging of electrical vehicles in which EDF, the French utility for electricity production and supply, is strongly involved. Once appropriately modeled, these problems appear as classical NP-hard graph theory problems, Max-k-Cut and Maximum Independent Set (MIS) respectively. To solve these problems, we develop dedicated extensions of the Quantum Approximate Optimization Algorithm (QAOA) \,\cite{Farhi14,Pichler18}. Besides providing numerical results evaluating the performances of these approaches on real data sets of electrical vehicle loads, the aim of this paper is to illustrate practical issues faced when trying to address real industrial problems with available NISQ frameworks. In particular, achieving good performances with quantum approaches requires the design of hardware-efficient procedures, that exploit the strengths of a given quantum processor. In this framework, hardware and software are jointly developed in order to optimize the execution of the overall implementation. Among the variety of platforms that are currently being investigated, programmable arrays of single neutral atoms manipulated by light beams appear as a very powerful and scalable technology to manipulate up to a few thousands qubits\,\cite{saffman_quantum_2010,Saffman16,Henriet20}. For the problems under consideration, we provide some implementation details on such platform as developed and commercialized by the company Pasqal.\\

The paper is organized as follows : section \ref{SCPbs} describes the smart-charging of electrical vehicles problems (SC1) and (SC2) which serve as use-cases  ; section \ref{QMAXKCUT} and \ref{QMIS} present the extension of QAOA to Max-k-Cut and the implementation of the approach proposed in Ref.\,\cite{Pichler18} to solve MIS, respectively ; section \ref{Results} is dedicated to our results and section \ref{future} to conclusions and further works. Complementary material is presented in the appendix.

\section{Two smart-charging problems and their modeling as NP-hard problems}\label{SCPbs}
It is admitted that in the next decades Electric Mobility will play a key role to solve major environmental and public health problems, thanks to the reduction of greenhouse gas and fine particles emissions it allows. Smart-charging appears as a mandatory condition to allow electric mobility expansion. Besides, the use of vehicle batteries both as energy storage and power supply devices ("Vehicle to Grid" or "V2G") could significantly improve the flexibility of the electric system, reducing high-peak of electricity demand thus saving significant energy.

Many difficult  problems lie behind this scheme in order to optimize the management of the electric system in terms of cost while satisfying various hard technical constraints. These include, among others, the modulation of electricity demand taking into account the potentially high demand specific to electric vehicle loads, the needs of electric vehicle users, the charging/discharging cycles of vehicle batteries and the reserves required to guarantee the frequency stability of the grid.\\

A great deal of theses problems takes the form of typical \textit{scheduling} and \textit{Operational Research} problems. They are large sized combinatorial optimization problems, many of them known to be NP-hard/complete.\\

The following sections describe two samples of these problems, and their modeling for quantum resolution.

\subsection{Vocabulary and common hypotheses}
Both problems will be tackled under the following assumptions:
\begin{itemize}
\item A load station is made up of several charging points, each of them loading at most a single electric vehicle (EV) at a given time step;
\item The charging points are \textit{parallel identical} machines that supply the same power. The charging time of a given EV is thus independent of the charging point it is scheduled on;
\item We consider neither additional job characteristics and constraints (release/due dates, charging profile imposed by the battery state) nor global resource constraints on the load station (maximal power deliverable at a given time step);
\item Preemption is not allowed: a load task cannot be interrupted to be resumed later, on the same charging point or another one.
\end{itemize}

\subsection{Minimization of Total Weighted Load Completion Time (SC1) and Max-k-Cut}
We consider $J = \lbrace 1, \dots ,n \rbrace$ charging jobs of $n$ EVs with durations $T = \lbrace t_1, \dots ,t_n \rbrace$, to be scheduled on a set $I = \lbrace 1, \dots,k \rbrace$ of $k$ charging points. An integer weight $w_j > 0$ is associated to each job $j$, measuring its importance. For example, we want to prioritize the charge of safety-related intervention vehicles. The time at which a load $j$ ends, called the \textit{completion time} is noted $C_j$ and we want to minimize the weighted total time of completion of the charges ~$\sum_{j\in J}w_jC_j$.\\

(SC1) is a classical scheduling problem known to be NP-hard in the general case, polynomial on a single machine or without priorities/weights attached to the jobs \cite{Graham79}. If the number of machines $m$ is fixed, (SC1) is NP-hard in the weak sense that it can be solved by pseudo-polynomial algorithms\footnote{An algorithm is said to be \textit{pseudo-polynomial} if it is  polynomial in the numeric values of its data, but super-polynomial in the length of their binary encoding.}, typically based on dynamic programming. In the case where $m$ is not fixed, (SC1) is NP-Hard in the strong sense, meaning that no such pseudo-polynomial algorithm exists except if P=NP.

Different approximation approaches have been studied for (SC1), until some PTAS in the general case and FPTAS\footnote{A \textit{Polynomial Time Approximation Scheme} (PTAS) is a family of $\varepsilon$-parametrized polynomial algorithms allowing to approximate optimal solutions "arbitrary closely" by a factor of $(1 - \varepsilon)$, $\varepsilon >0$. When the algorithms are also polynomial in their parameter $\varepsilon$, one talks of \textit{Full Polynomial Time Approximation Scheme} (FPTAS)} in the weakly NP-Hard variant were established \cite{Skutella98, Yang03, Skutella99, Woeginger00}.\\
Interestingly for our purpose, some of these approximation approaches were based on the reformulation of (SC1) as a weighted Max-k-Cut problem, a problem that can be tackled by the Quantum Approximation Optimization Algorithm (QAOA), at least in its $k=2$ set up i.e.~Max-Cut. \\ 

Consider the complete graph $G=(V,E)$ whose vertices $V$ correspond to the $n$ jobs in $J$, and with a weight assigned to each edge $(k,j)$ in $E$ defined by $w_{kj}= \min\{w_k t_j ; w_j t_k\}$. The maximal $k$-cut of this graph provides with the optimal affectation of the $n$ jobs on the $k$ machines. Informally, this relies on the well known "Smith Rule"\,\cite{Smith56} which states that once jobs have been affected to machines, the optimal scheduling is given by executing them in a non-increasing order defined by the ratio $\frac{w_j}{t_j}$. On this basis, minimizing the weighted total completion time of the tasks can be shown to be equivalent to maximizing the above weight on crossing edges between the subsets of a $m$-partition of $V$\,\cite{Skutella98, Yang03}.\\

For both Max-Cut and Max-k-Cut, known polynomial randomized approximation algorithms obtain high \textit{approximation ratios} $C/C_{\text{opt}}$ -- where $C$ represents the average value of the solution provided by the algorithm and $C_{\text{opt}}$ the optimal one --, namely $0.878567$ \cite{Goemans95} and $\left(1-\frac{1}{k}+ 2\frac{\ln k}{k^2}\right)$ \cite{Frieze95}, respectively (the latter is improved for small values of $m$ in Ref.\,\cite{Klerk04}). Moreover, improving these ratios is proved to be NP-Hard, unless some highly-believed conjectures of Complexity Theory would be false. In the case of Max-Cut, improving the approximation ratio from $0.878567$ up to $(\frac{16}{17}=0.941176)$ is NP-Hard \cite{Hastad01} and thus out of reach of any polynomial classical algorithms unless $P=NP$, as it is to increase the ratio upon $1-1/34 k$ for Max-k-Cut \cite{Kann97}.
Likewise, both problems are APX-Hard and thus, unless $P=NP$, they have no PTAS \cite {Papadimitriou91,Frieze95}.
Finally, tighter results were proven under the "Unique Games Conjecture", namely that improving the above original ratios for Max-Cut and Max-k-Cut established in Ref.\,\cite{Goemans95} and Ref.\,\cite{Frieze95} is NP-Hard \cite{Khot07}.\\

This means that such improvements by some quantum algorithms would establish a "quantum supremacy" on NP-Complete problems, unless these conjectures turn out to be false, which is considered unlikely. Here, we do not aim at achieving such an approximation ratio improvement with quantum algorithms in the general/worst-case. Alternately, we analyze their performances on some graphs drawn from real-world problems considering the performances obtained by the best-known randomized approximation classical algorithms. We observe that QAOA outperforms Goemans and Williamson ratio on these particular instances of Max-Cut.  Similar results were observed in Ref.\,\cite{Crooks18} on small random graphs. Establishing that such improvement would hold true for worst-case and large graphs instances is still an open question, and will remain so as long as we stay in the NISQ era. In any case, reaching similar approximation ratios as classical solutions, but faster or at a lower energy cost would already be of significant value for industries.\\

In this context, it should be noted that although these limits to improving the approximation ratios are valid for general/worst-case instances of Max-k-Cut, it does not mean that they are valid for particular instances with a specific structure. Thus, since efficient classical approximation schemes are known for (SC1), we can expect its instances reformulated as instances of Max-k-Cut to be easier to approximate than general/worst-case ones. We will see that this is indeed the case, both when using classical and quantum algorithms.

\subsection{Optimal Scheduling of Load Time Intervals within Groups (SC2) and MIS}\label{LIGMIS}
We now consider the following problem (SC2): given a set of load tasks represented as \textit{intervals} on a timeline, such that each of them belongs to a specific \textit{group} , for example distinct vehicle fleets of a company, select a subset of these loads (i) which maximizes the number of non-overlapping tasks and (ii) such that at most one load in each group is completed. The goal is here to both minimize the completion time of the selected loads and to guarantee that no group will be over-represented in the schedule.
 
This problem belongs to the class of \textit{Interval Scheduling} problems \cite{Kolen07}. More precisely, it is a \textit{Group Interval Scheduling}, or \textit{Job Interval Selection} problem. It can be restricted without loss of generality to the case where all the groups contain the same number of tasks, $k$. It is NP-Complete for $k \geq 3$, and has no PTAS for $k \geq 2$ unless $P=NP$ \cite{Spieksma99}. Some polynomial approximation ratios have been obtained in the general case, namely $0.5$ in Ref.\,\cite{Spieksma99}, improved to $0.63211$ in Ref.\,\cite{Chuzhoy06}, while polynomial algorithms exist in cases where some parameters are fixed \cite{Bevern14}.\\

Let $I = \lbrace (s_1,e_1) \dots (s_{nk},e_{nk}) \rbrace$, where $n$ is the number of groups, be the set of intervals representing load job starting and ending dates, and $G=(V,E)$ be the graph whose vertices in $V$ correspond to intervals in $I$, and with an edge $(i,j)$ in $E$ \textit{iff} interval $i$ and $j$ overlap \textit{\textbf{or}} $i$ and $j$ belong to the same group. Clearly, an \textit{independent set} of this graph, i.e.~a set of vertices no two of which are adjacent, represent a feasible solution of the problem, and its Maximum Independent Set (MIS) is the optimal one\footnote{The above formulation supposes that a \textit{starting date} is affected to each load task, in order to represent it as an interval on the time line. In a real smart-charging management system, such dates could be fixed by the users of the vehicles, imposed by technical constraints or decided by the smart-charging manager.}.\\

Following our protocol for quantum algorithm qualification, we will limit ourselves to specific instances of (SC2) that can be formulated as MIS on two-dimensional \textit{Unit-Disk} (UD) graphs. These geometrical graph are graphs in which two vertices are coupled by an edge if the distance between them is below a threshold value. This choice is motivated by two reasons. First, as described at section \ref{QMIS} below, Pasqal's neutral atom quantum processor is particularly well suited to natively implement the MIS on Unit-Disk graphs. Second, the MIS on such graphs, although remaining NP-Complete, is known to be "less difficult" to approximate than the MIS on general graphs, and has a PTAS while the general MIS does not~\cite{Chan02,Nieberg04}. Of course, this protocol requires to transform (SC2) graphs to Unit-Disk graphs, a procedure we discussed in section \ref{FromIG2UDG}.

\section{Quantum approaches to smart-charging problems}
In this section, we describe how to use quantum approaches for solving the problems presented above. 
\subsection{QAOA in a nutshell}
The "Quantum Approximate Optimization Algorithm" (QAOA) computes approximate solutions to combinatorial optimization problems, with a theoretical guarantee of convergence when the depth of the quantum circuit increases~\cite{Farhi14}.\\

\begin{figure}[h!]
    \centering
    \includegraphics[width=0.8\textwidth]{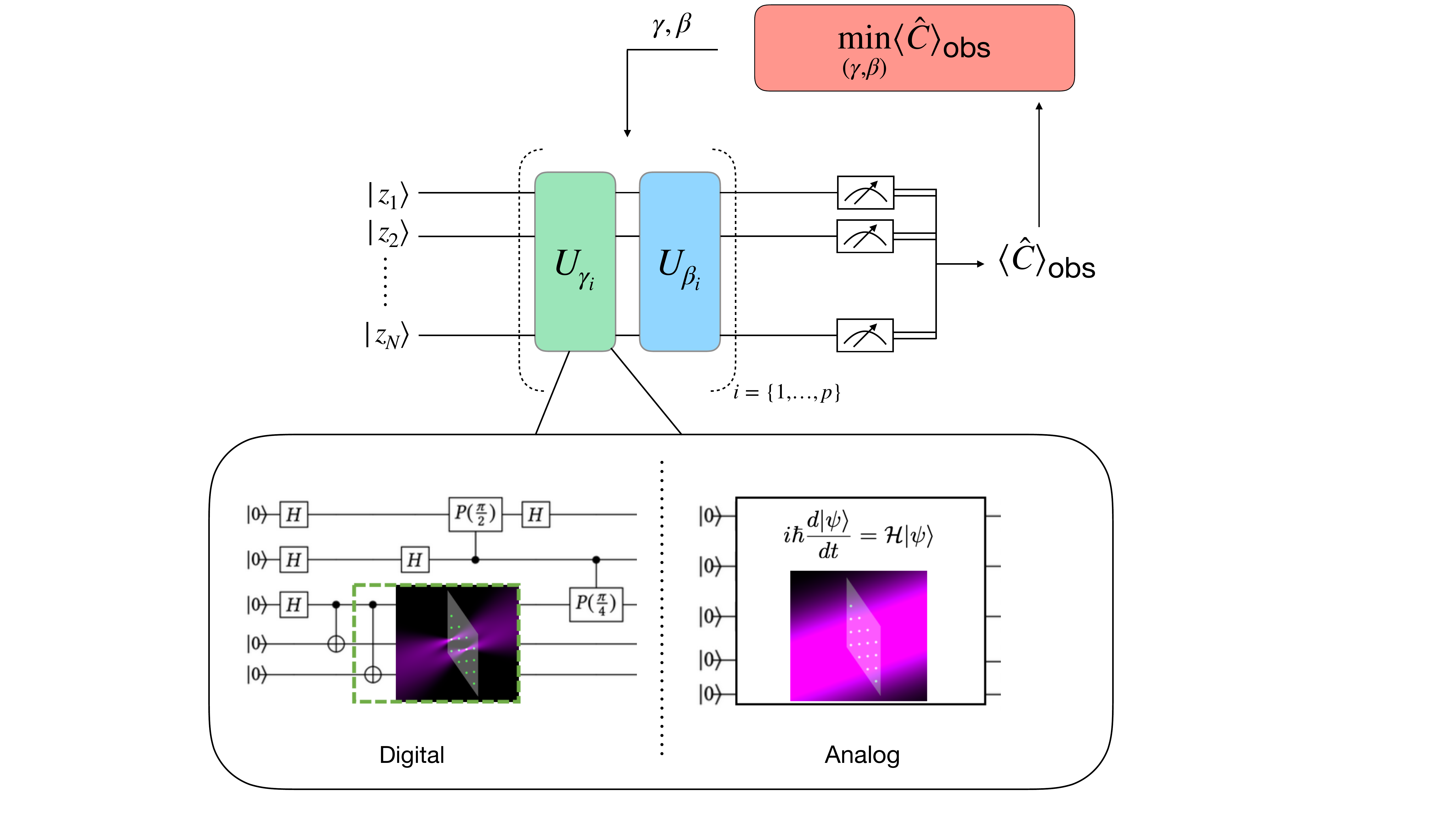}
    \caption{Principle of the QAOA algorithm. A quantum processor, which can be operated in either digital or analog mode, is used to prepare an ansatz wavefunction from which we construct the mean value $\langle \hat{C} \rangle_{\text{obs}}$ using numerous measurements. A classical optimizer then updates the variational parameters. Some problems are naturally tailored to the analog mode of the platform, while others require a digital mapping. In digital mode, the unitaries are built from quantum circuits, made of elementary quantum gates acting each on one or a few qubits. In the analog mode, the unitaries are built from sequences of Hamiltonians that can be controlled in a continuous manner.
 }
    \label{fig:qaoa}
\end{figure}

QAOA is a variational algorithm for combinatorial problems in which a quantum processor works hand-in-hand with a classical counterpart, as illustrated in Fig.\,\ref{fig:qaoa} (see Ref.\,\cite{Cerezo20} for a review on variational algorithm). The quantum processor is used to prepare a wave function $|\bm{z}_{\bm{\gamma},\bm{\beta}}\rangle$. In the most general case, $|\bm{z}\rangle=|z_1 z_2 ... z_n\rangle$ represents a $n$-qudit state vector, with $z_i \in \{0,1,..,d\}$, and the subscript in $|\bm{z}_{\bm{\gamma},\bm{\beta}}\rangle$ indicates that the state belongs to a family of states that is parameterized by the angles ${\bm{\gamma}}$ and ${\bm{\beta}}$. More specifically, $|\bm{z}_{\bm{\gamma},\bm{\beta}}\rangle$ is generated by the successive application of unitaries generated by the non-commutative operators $\hat{M}$ and $\hat{C}$, with angles given by $\bm{\beta}=(\beta_1,\beta_2,...,\beta_p)$ and $\bm{\gamma}=(\gamma_1,\gamma_2,. ..,\gamma_p)$, respectively. Given an initial state $|\bm{z}_0 \rangle$, the wavefunction prepared by the quantum processor takes the following form, \begin{equation}
|\bm{z}_{\bm{\gamma},\bm{\beta}}\rangle = e^{-i\beta_p \hat{M}}e^{-i\gamma_p \hat{C}}\dots e^{-i\beta_1 \hat{M}}e^{-i\gamma_1  \hat{C}}|\bm{z}_0 \rangle.\label{eq:ansatz}
\end{equation}
A common choice for the cost operator $\hat{C}$ (also sometimes referred to as energy operator) is the diagonal operator in the computational basis, $\hat{C}|\bm{z}\rangle = C(\bm{z})|\bm{z}\rangle$, where $C(.)$ is the cost function to be optimized for, while the mixing operator $\hat{M}$ induces transitions between states in the computational basis\,\cite{Farhi14,Hadfield2017}. In the following, we will always reformulate our problems under the form of minimization problems, by changing the sign of the cost function of the original problems. The ground state of the energy operator $\hat{C}$ corresponds to the optimal solution to the optimization problem. The dimension $p$ of the vectors $\bm{\gamma}$ and $\bm{\beta}$ is called the depth of the algorithm. 

The quantum state is then measured to construct an statistical estimator $\langle \hat{C} \rangle_{\text{obs}}$ of the cost function to be minimized. A classical optimization procedure uses this estimator to update the variational parameters $\bm{\gamma}$ and $\bm{\beta} $ for the next iteration. This loop repeats until convergence to a final state, from which an estimate of the solution to the problem is extracted. The approximation ratio $\langle \hat{C} \rangle/C_{\text{opt}}$, where $C_{\text{opt}}=\min_{\textbf{z}} C(\textbf{z})$, measures the quality of the approximation yielded by QAOA.\\

In the following, we will consider two distinct ways to prepare the ansatz wavefunction (\ref{eq:ansatz}) on Pasqal quantum processors. In the first one, the quantum processor is used in digital mode, and the trial wavefunction is the output of a quantum circuit composed of discrete quantum gates. In the second one, the quantum processor is used in an analog manner and the trial wavefunction results from the application of a continuously parameterized Hamiltonian. \\
 
\subsection{Max-k-Cut}\label{QMAXKCUT}

We describe here an extension of QAOA to deal with weighted Max-k-Cut problems and using quantum circuits. We also introduce a hardware-efficient implementation of this procedure on arrays of neutral atoms.

\subsubsection{Encoding on the quantum processor}

We show here how to extend QAOA to Max-k-Cut problems on weighted graphs. Denoting by $z_i\in \{1,2,..,k\}$ the subset node $i$ belongs to, the goal of Max-k-Cut is to minimize the cost function: 
\begin{equation}
    C(z_1,\dots,z_N)=-\sum_{u = 1}^N \sum_{v=u}^N w_{uv}(1-\delta_{z_u,z_v}),
    \label{eq:classic_cost}
\end{equation}
where the edge $w_{uv}$ participates to the cost function if nodes $u$ and $v$ are in separated subsets.\\

Originally QAOA was designed for optimization problems of type QUBO (Quadratic unconstrained binary optimization). In that case,  a problem with $n$ variables naturally maps to a system of $n$ qubits. For optimization over integer-valued variables one has to define a correspondence between basis states of a Hilbert space of some dimension and solutions to the initial problem, and then restrict the quantum evolution to the feasible domain.\\

A possible approach to tackle this issue is to apply the unary encoding \cite{Hadfield2017} that for any natural $k > 2$ attributes $k$ qubits to each node and a node $u$ is colored in $c$ if $|z_u\rangle = |0\dots010\dots0\rangle$ where the bitstring $z_u$ has only one $1$ on the position $c$. This encoding uses $Nk$ qubits for an instance with $N$ nodes and requires some non-trivial modifications to the mixing operator $M$ in order to keep the evolution in the feasible subspace, which is spanned by basis states with only one $1$ per node. This encoding was also used in Ref.\,\cite{Kudo2018} for solving graph-coloring problem with \textit{QAA}.\\

We suggest instead to use the conventional binary encoding of integers for $k=2^l$. In this encoding each node is associated to a set of $l$ qubits $\mathbf{z} = \{ z^{(0)}, \dots,z^{(l-1)}\}$ that will indicate the color of the node. More specifically, having qubits in the state $\mathbf{z}_u$ corresponds to having the node $u$ in the subset $z_u=z^{(0)}z^{(1)}...z^{(l-1)}$. Note that we need $Nl \approx N\ln k$ qubits to encode the entire coloring of the graph. In the specific case where $k=2^l$, the computational basis that spans the Hilbert space of our platform corresponds exactly to all the $2^{Nl}$ possible colorings of the graph. It means that, contrary to the unary encoding, no modification to the operator $M$ is required. This approach is easier to compile to elementary gates and less expensive in terms of number of qubits than the unary encoding. A natural extension for all natural $k$ would be to use qudits instead of qubits \,\cite{Bravyi20}.\\

\begin{figure}[h!]

\begin{equation*}
    \Qcircuit @C=1.0em @R=0.0em @!R {
                \lstick{ {q}_{u}^{(0)} : } & \ctrl{2} & \qw & \ctrl{2} & \qw & \qw & \qw& \ctrl{1} & \qw & \qw & \qw & \qw & \qw & \ctrl{1} \\
                \lstick{ {q}_{u}^{(1)} : } & \qw & \qw & \qw & \ctrl{2} & \qw & \ctrl{2} & \targ & \ctrl{1} & \qw & \qw & \qw & \ctrl{1} & \targ\\
                \lstick{ {q}_{v}^{(0)} : } & \targ & \gate{R_z(\frac{w_{uv} \times \gamma}{2})} & \targ & \qw & \qw & \qw & \qw & \targ & \ctrl{1} & \qw & \ctrl{1} & \targ & \qw \\
                \lstick{ {q}_{v}^{(1)} : } & \qw & \qw & \qw & \targ & \gate{R_z(\frac{w_{uv} \times \gamma}{2})} & \targ& \qw & \qw & \targ & \gate{R_z(\frac{w_{uv} \times \gamma}{2})} & \targ & \qw  & \qw \\
         }
\end{equation*}
\caption{ A quantum circuit implementing a term $e^{-i\gamma C_{u,v}}$ corresponding to the edge $\langle u,v \rangle$ in binary encoding for $k=4$}
\label{fig:circ_for_hc}
\end{figure}
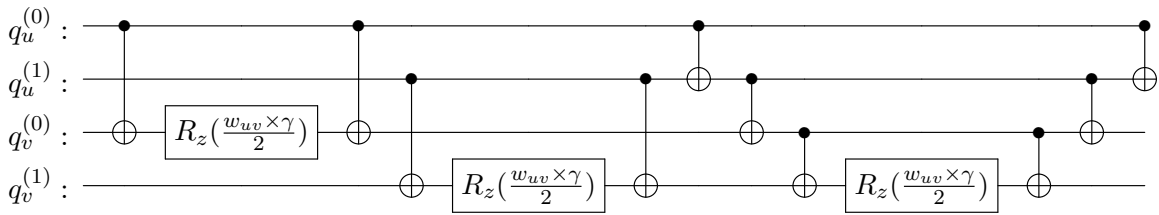

For binary encoding we can build a cost operator $\hat{C}$ that is diagonal in the computational basis, such that $\hat{C}|\mathbf{z}_1 \dots  \mathbf{z}_N \rangle = C(\mathbf{z}_1 \dots  \mathbf{z}_N )|\mathbf{z}_1 \dots  \mathbf{z}_N \rangle,$ by writing :

\begin{equation}
    \hat{C} = - \sum_{1 \leq u < v \leq N}  w_{uv} \left (1 - \frac{1 + \sigma_u^{(0)}\sigma_v^{(0)}}{2} \dots \frac{1 + \sigma_u^{(l-1)}\sigma_v^{(l-1)}}{2} \right),
    \label{eq:H_c}
\end{equation}
where $\sigma_u^{(i)}$ corresponds to a Pauli-Z matrix $\sigma_z$ acting on the atom $i \in \{0, \dots,l-1\}$ associated to node $u$. The operator $C$ in Eq. (\ref{eq:H_c}) can be decomposed as a sum of operators with $\{2,4, \dots, 2l \}$-body interaction terms. Interaction terms involving more than 2-body operators are not directly implementable on most quantum computing platforms which only support 2-qubit gates. Instead, they can be decomposed as sums of two-body terms, which can be realized with CNOT gates. An example of the resulting circuit for an edge is shown in Fig. \ref{fig:circ_for_hc}.

Because of the completeness of the Max-k-Cut graphs that we are studying, the cost operator in Eq. (\ref{eq:H_c}) involves coupling terms between all qubits. However, all quantum chips have a finite connectivity in practice. The physical realization of the desired terms between remote qubits thus requires the introduction of a large number of SWAP gates, that are detrimental to the performance of the procedure. Next, we introduce a hardware-efficient implementation of Max-k-Cut on Rydberg atom arrays which minimizes this overhead. 

\subsubsection{Towards a hardware-efficient implementation of Max-k-Cut on Rydberg atom arrays}

Mapping the QUBO representation of the Max-k-Cut problem as discussed above introduces on the one hand an overhead in number of CNOT gates as a result of the decomposition of multi-spin terms into pair interactions. An alternative to the QUBO representation of the problem is the parity encoding, originally proposed in Ref. \cite{lechner2015quantum} (Lechner-Hauke-Zoller scheme) and recently extended to applications using QAOA \cite{lechner2020quantum}. \\

Let us summarize here the main steps of the parity transformation for completeness and then apply it to the Max-k-Cut problem. The parity encoding encodes the relative orientation of several logical qubits as a single physical qubit.  With this encoding, each physical qubit represents the parity of several logical qubits, e.g.~the physical spin $\hat{z}_{12} = z_1 z_2$ represents the product of two logical bits in the QUBO representation. Here, $\hat{z}_{ij}$ is used as short notation of the z-component of the physical qubit $\sigma_{ij}^{(z)}$ and $z_i$ is used for the the logical qubits $\sigma_{i}^{(z)}$. The parity encoding represents the full problem in terms of the parity variables. This is not trivially possible, as the number of degrees of freedom of the logical graph are $N$ (the  number of logical spins), while the number of physical spins is $K$, the number of edges in the graph. To compensate for this difference, a number of $C = K - N +1$ constraints need to be introduced. These constraints are conditions on the physical qubits that have to be fulfilled. In the Lechner-Hauke-Zoller scheme, these constraints are constructed from closed loops in the logical graph which results in a physical implementation that consists  of $K$ qubits and $K-N+1$ 4-body constraints. Remarkably, the problem is fully encoded in the local fields while all interactions are uniform and problem independent \cite{lechner2015quantum}. 

\begin{figure}[h!]
    \centering
    \includegraphics[width=0.65\textwidth]{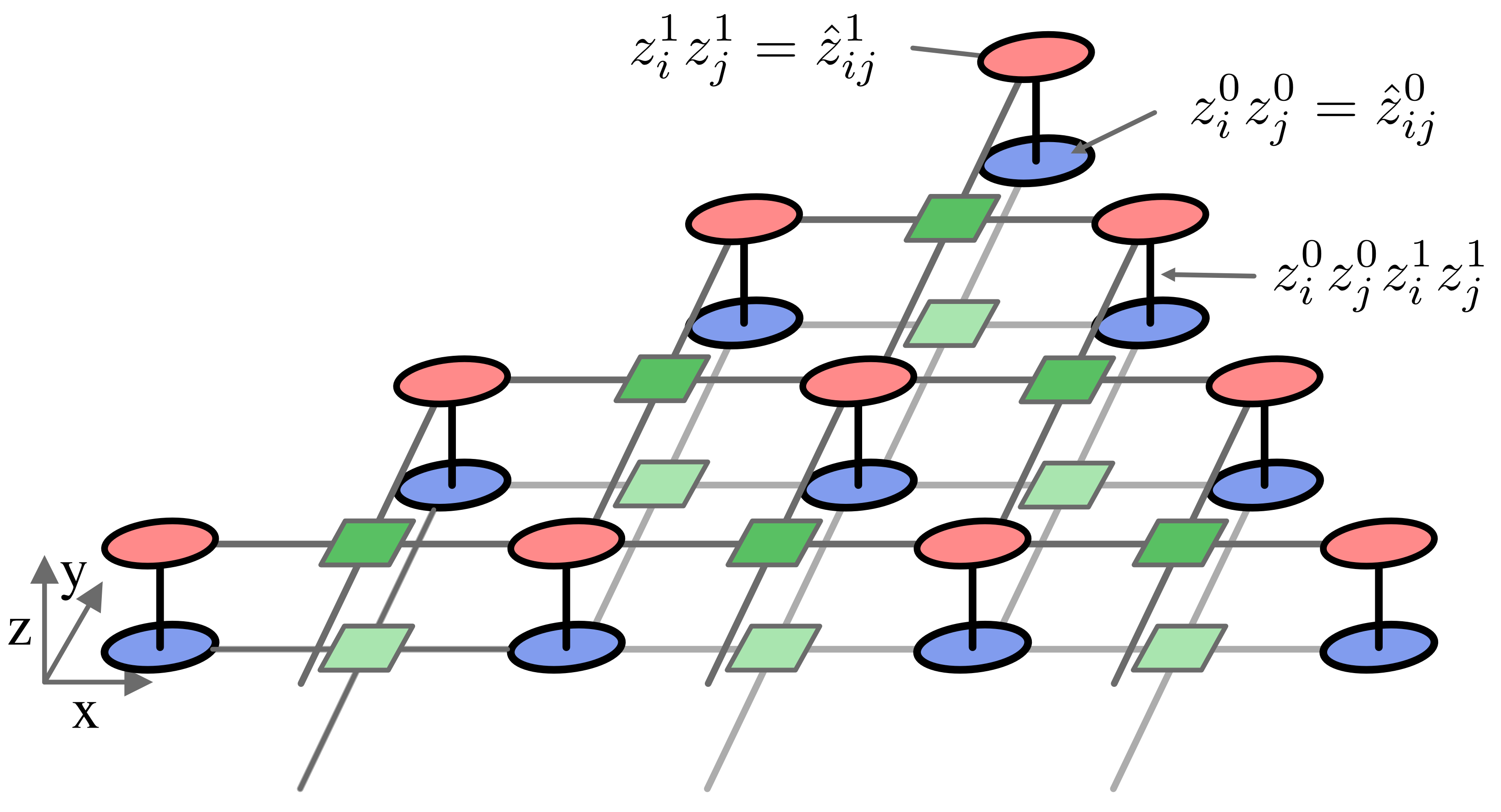}
    \caption{Parity transformation of the Max-4-cut problem. The setup consists of two layers of qubits (blue bottom and red top). The bottom layer represents the 0th bit of the color and the top layer the 1st bit of the color. The qubits within the same layer interact via 4-body constraints \cite{lechner2015quantum} (green squares). The interaction terms  $\hat{z}^0_{i}\hat{z}^1_{i} = z^0_i z^0_j z^1_i z^1_j$ are efficiently encoded as pair gates between neighboring qubits of the two layers (black lines). Note, that the 4-body constraints are decomposed into parallelizable CNOT gates within the layers \cite{lechner2020quantum}. }
    \label{fig:fclhz}
\end{figure}

Applying the parity encoding to the Max-4-cut problem allows us to reduce the number of CNOT gates to order $N^2$. In the first step, we make use of the parity transformation of each term $z^0_i z^0_j \rightarrow \hat{z}^0_{ij}$ and $z^1_i z^1_j \rightarrow \hat{z}^1_{ij}$. As a second step we transform the 4-body term of the two-coloring problem which has the form $z^0_i z^0_j z^1_i z^1_j$. We notice, that this term in the parity encoding is $\hat{z}^0_{i}\hat{z}^1_{i}$ and thus a simple pair interaction between two physical qubit. Remarkably, the pair interaction can be performed between nearest neighbors by introducing a two-layer setup as depicted in Fig. \ref{fig:fclhz}. \\

The cost function using the above construction reads as  
\begin{equation}
	C = \sum_{i} \left( \hat{z}^1_{i} +  \hat{z}^0_{i} + \hat{z}^0_{i}\hat{z}^1_{i} \right) + \sum_l C_4 \hat{z}^0_{i}\hat{z}^0_{j}\hat{z}^0_{k}\hat{z}^0_{l}
\end{equation}\

Let us now comment on the resources requirements for this hardware-efficient implementation. The number of qubits is $N(N-1)$ and the number of CNOT gates is $9N^2 - 41N + 48$. In the above example for $N=5$ the number of qubits is $N_p=10$ and number of CNOT Gates is $N_c = 68$. In addition, the CNOT gates can be fully parallelized which results in a constant depth of global gates. In comparison, the number of qubits in the standard gate model is $N$ and the number of CNOT gates to be calculated but assumed to be order $N^3$. This encoding provides thus a strong advantage in platforms which show a capacity to scale up the number of qubits.

With a number of thousands of qubits of next generation neutral atom devices\,\cite{Henriet20} in combination with the prospect of full parallelizability of the parity encoding, Max-k-Cut represents a realistic application of near term quantum devices.

\subsection{MIS}\label{QMIS}
In this Section we present how to solve a MIS problem using a QAOA procedure on a neutral atom processor working in an analog mode\,\cite{Pichler18,Henriet20a}. We first show how an analog control over the atoms allows us to realize a Hamiltonian reproducing the MIS cost function in the case of Unit Disks (UD) graphs. Then, we discuss a procedure enabling us to transform more generic interval graphs to UD graphs.

\subsubsection{Rydberg blockade and graph independent sets}

When looking for the MIS of a graph, we separate the nodes into two distinct classes: an independence one and the others. We can attribute a status $z$ to each node, where $z_i = 1$  if node $i$ is attributed to the independent set, and $z_i=0$ otherwise. The Maximum Independent Set corresponds to  the minima of the following cost function: 

\begin{equation}
   C(z_1,\dots,z_N) = -\sum_{i=1}^N z_i + U \sum_{\langle i,j \rangle}z_i z_j
 \label{cost_function}
\end{equation}
where $U \gg 1$  and $\langle i,j \rangle$ represents adjacent nodes (\textit{i.e.} there is a link between node $i$ and $j$). In this cost function, we want to promote a maximal number of atoms to the $1$ state, but the fact that $U \gg 1$  strongly penalizes two adjacent vertices in state 1. The minimum of $C(z_0,\dots,z_N)$ therefore corresponds to the maximum independent set of the graph.\\

Interestingly, the operator $\hat{C}$ associated with the cost function of Eq. (\ref{cost_function}) can be natively realized on a neutral atom platform\,\cite{Pichler18}, with some constraints on the graph edges. We map a  ground state and a Rydberg state of each atom to a spin $1/2$, where $|1 \rangle = |r \rangle$ is a Rydberg state and $|0 \rangle = |g \rangle$ is a ground state. An atom in a Rydberg state has an excited electron with a very high principal quantum number and therefore exhibits a huge electric dipole moment. As such, when two atoms are excited to Rydberg states, they exhibit a strong van der Waals interaction. Placing $N$ atoms at positions $\textbf{r}_j$ in a 2D plane, and coupling the ground state $|0\rangle$ to the Rydberg state $|1\rangle$ with a laser system enables the realization of the Hamiltonian :
\begin{equation}
    H= \sum_{i=1}^N \frac{\hbar\Omega}{2} \sigma_i^x - \sum_{i=1}^N \frac{\hbar \delta}{2}  \sigma_i^z+\sum_{j<i}\frac{C_6}{|\textbf{r}_i-\textbf{r}_j|^{6}} n_i n_j.
\label{eq:ising_Hamiltonian}
\end{equation}
Here, $\Omega$ and $\delta$ are respectively the Rabi frequency and detuning of the laser system and $\hbar$ is the reduced Planck constant. The first two terms of Eq. (\ref{eq:ising_Hamiltonian}) govern the transition between states $|0\rangle$ and $|1 \rangle$ induced by the laser, while the third term represents the repulsive Van der Waals interaction between atoms in the $|1\rangle$ state. More precisely, $n_i = (\sigma_i
^z + 1)/2$ counts the number of Rydberg excitations at position $i$. The interaction strength between two atoms decays as $|\textbf{r}_i-\textbf{r}_j|^{-6}$. \\

\begin{figure}[h!]
    \centering
    \includegraphics[width=0.8\textwidth]{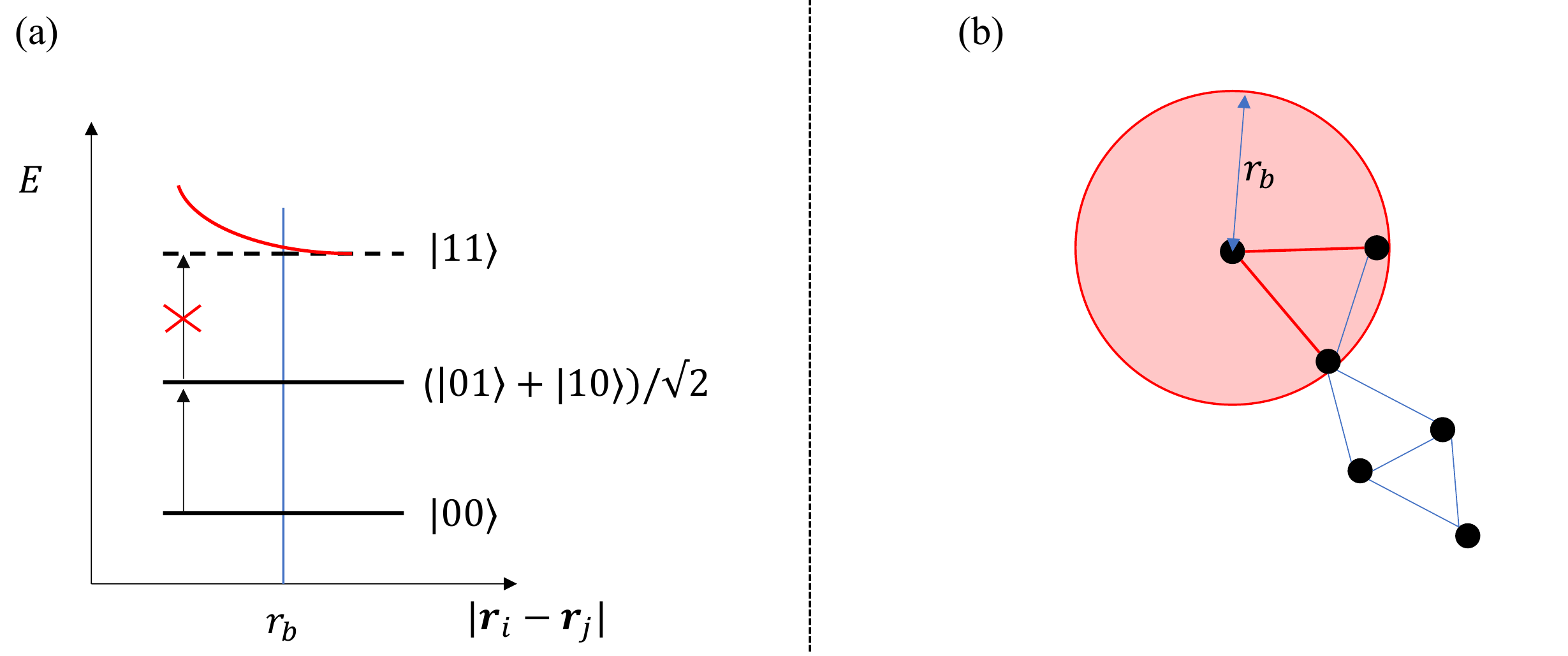}
    \caption{(a) Illustration of the Rydberg blockade effect. When two atoms are far apart, $|\textbf{r}_i-\textbf{r}_j|>r_b$, they don't interact. On the other hand, if they are separated by less than the Rydberg blockade radius, $|\textbf{r}_i-\textbf{r}_j|<r_b$, a strong interaction prevents the two atoms to be in the state $|1\rangle$ at the same time. (b) Rydberg blockade and independent sets of a graph. Rydberg atoms correspond to the nodes of a UD-graph. There are edges between adjacent nodes if the distance between them is smaller than the Rydberg blockade radius $r_b$, as illustrated for the top-left node. Due to the Rydberg blockade effect, the dynamics of the system is restricted to independent sets.   }
    \label{fig:rydberg_blockade}
\end{figure}

The shift in energy originating from the presence of two nearby excited atoms induces the so-called \textit{Rydberg blockade} phenomenon, illustrated in Fig.\,\ref{fig:rydberg_blockade}(a). More precisely, if two atoms are separated by a distance smaller than the Rydberg blockade radius $r_b = (C_6/\hbar \Omega)^{1/6}$, the repulsive interaction will prevent them from being excited at the same time. On the other hand, the sharp decay of the interaction allows us to neglect this interaction term for atoms distant of more than $r_b$. As such, for $\Omega=0$, the Hamiltonian in Eq. (\ref{eq:ising_Hamiltonian}) is diagonal in the computational basis and enables to realize $H |z_1,\dots,z_N\rangle=(\hbar \delta/2) C(z_1,\dots,z_N)|z_1,\dots,z_N\rangle$, with the cost function specified in Eq. (\ref{cost_function}), and for which there is a link between atoms $i$ and $j$ if they are closer than $r_b$ apart.\\

We have seen that atom arrays enable to study UD-MIS problems with QAOA using an analog control, which will likely offer better performances as compared to a digital approach. In the following, we present a way to transform interval graphs into UD graphs.

\subsubsection{From Interval Graphs to Unit Disk Graphs}\label{FromIG2UDG}

Graphs such as the ones used to model our Job Interval Selection problem (SC2) do not correspond to UD graphs, as they are unions of \textit{interval intersection graphs} (the edges corresponding to the overlapping tasks) and of \textit{cluster graphs} (the set of complete disjoint cliques corresponding to the groups) and thus \textit{one-dimensional} intersection graphs, rather than \textit{two-dimensional} ones. Hence, we have to transform our scheduling graphs in order to implement the search of their MIS on the quantum  machine.\\

The correspondence between the original problem graph and the locations of the Rydberg atom/qubits in the machine can be obtained by solving the following continuous quadratic problem (UD):
\begin{flalign}
&Min_{x_q,y_q}\lbrace L \rbrace\\
&s.t.\nonumber\\
&\forall \langle i,j \rangle \in E,  \rho^2 \le (x_j - x_i)^2 + (y_j - y_i)^2 \le r^2\label{CtrDist}\\
&\forall \langle i,j \rangle \in \overline{E},  (x_j - x_i)^2 + (y_j - y_i)^2 > r^2\label{CtrDistComp}\\
&\forall i \in V, x_i < L; y_i < L\\
& L \leq \overline{L}
\end{flalign}
where:
\begin{itemize}
\item $G=(V,E)$ is the original interval graph of the problem;
\item $\overline{G}=(V,\overline{E})$ is $G$'s complementary, i.e.~the graph whose vertices $(i,j)$ are connected iff they are not connected in $G$;
\item $r$ is the Rydberg blockade radius, i.e.~the upper bound on the distance between two connected vertices, $\rho\sim r/3$  a given factor defining the lower bound on their distance, illustrating the minimal spacing between separated atoms that is experimentally feasible;
\item $(x_i, y_i)$ are the coordinates in the euclidean plan of the atom/qubit representing the node $i\in V$;
\item $L$ is the minimum length of the side of the square  required to embedded the graph while enforcing constraints above;
\item $\overline{L}$ defines the maximal surface available to place the atoms on the machine. 
\end{itemize}

Experimentally, the Rydberg blockade radius $r$ can be around 15$\mu$m, for minimal distances between atoms of the order of 5$\mu$m, and the atoms are contained in a region characterized by $\bar{L}\sim 100 \mu$m.\\

Due to the lower bound in constraints (\ref{CtrDist}) and to constraints (\ref{CtrDistComp}), this problem is not convex, and is thus NP-hard in the general case \cite{Park17,Elloumi19,Billionnet08}. Furthermore, $L$ clearly depends on the machine characteristics $(r,\rho,\overline{L})$, and on the structure of the original interval graph $G$, and the (UD) problem does not necessarily have feasible solutions. For example, it is clear that for poorly connected graphs, i.e with few load intervals overlapping or many groups with a low $k$, the requirement to keep away from each other atoms corresponding to non-connected nodes might render the problem infeasible in the square $\overline{L}\times\overline{L}$. On the contrary, strongly connected graphs should make difficult to satisfy the constraint of bringing together the atoms corresponding to connected nodes in the rings delimited by $(r,\rho)$. We have tested different methods to solve the (UD) problem, allowing to produce feasible locations of Rydberg qubits  on the quantum processor for various real smart-charging graphs in a short elapsed computation time. They are presented in Appendix \ref{AnnexQUD}.\\

However it is worth noting that we had to solve the (UD) problem only for graph instances whose vertices's number is limited. It is clear that tackling larger load scheduling problems on larger quantum processors with this method could turn out to be difficult, due to the cost of computing the corresponding "UnitDisk" graph of Rydberg atoms. Overcoming this mapping issue is an important research objective that we leave for future work.

\section{Results}\label{Results}

\subsection{Analytical result for Max-Cut: the mean cost in the QAOA$_1$ state}

An analytical expression for Max-Cut on unweighted graphs for $p=1$ was derived in \,\cite{Wang2018}. We extend this result to the case of complete weighted graphs. For an edge $\langle u, v \rangle$ the mean value of $C_{u,v}$ for chosen $\gamma, \beta$ is:


\begin{flalign}
\langle \mathbf{z}_{\gamma, \beta} |C_{u,v}| \mathbf{z}_{\gamma, \beta} \rangle = &\frac{w_{uv}}{2} \left( 1 +\frac{\sin(4\beta)\sin(\gamma w_{uv})}{2}
\left[\prod_{x \neq u, v}\cos(\gamma w_{ux}) + \prod_{x \neq u, v}\cos(\gamma w_{vx}) \right] + \right.\notag \\
& \left. - \frac{\sin^2(2\beta)}{2}\left[\prod_{x \neq u, v} \cos(\gamma (w_{ux} - w_{vx}))- \prod_{x \neq u, v} \cos(\gamma (w_{ux} + w_{vx}))\right] \right)
\label{eq:analytical_maxcut_depth_1}
\end{flalign}


This is an important result for numerical experiments as the obtained expression allows to compute the mean cost in the QAOA$_1$ state without any call to a (simulated) quantum computer in polynomial time, which in turn is useful to analyze the performance of QAOA. 

In this expression, we realize that there is a close link between the value of the QAOA angle $\gamma$ and the weight of the edge $w_{uv}$. Discussions on the normalization of the graph edges  and its impact on the energy landscape are covered in Appendix \ref{Param_opt}.

\subsection{Numerical results}

In the following, we present a performance analysis of the various procedures presented above. For Max-Cut at depth $p=1$, we use the analytical formula (\ref{eq:analytical_maxcut_depth_1}). For Max-4-Cut, MIS, and Max-Cut at depth $p>1$, we compute the mean cost by Monte-Carlo estimation, by simulating the quantum evolution either on the Atos Quantum Learning Machine~\cite{atos} or using the QuTIP library \cite{Johansson_2013} on the OCCIGEN supercomputer based in Montpellier, France.

\subsubsection{Real Data Set Used}
Data were driven from a set of $2250$ loads performed during May 2017 on identical charging points of the Belib's network of load stations located in Paris, France\,\cite{Belib17}.\\

For both problems (SC1) and (SC2), an instance is a series of chronological loads characterized by their duration for (SC1) and starting/end times for (SC2). The instance size is the number of loads it contains. A data-set is a set of instances whose first load is randomly chosen among the 2250, according to a uniform law. Once an instance is built:
\begin{itemize}
\item for (SC1), a priority is randomly affected to each load, according to a Poisson's law, enforcing a constant distribution of the different priority levels in each instance\footnote{The "brute" weights from the data yield high values in the cost Hamiltonian, leading to a highly fluctuating energy landscape for $QAOA_1$ (cf. equation \ref{eq:analytical_maxcut_depth_1}). A re-weighting was therefore done in order to obtain a smoother landscape for better optimization. (see Appendix \ref{Param_opt}).};
\item for (SC2), the belonging to a group is randomly affected to each load according two a uniform law parameterized by the number of groups and the number of loads in the instance.
\end{itemize}

\subsubsection{Minimization of Total Weighted Load Completion Time (SC1)}\label{QMAXRes}

First, we compare the performance of QAOA and the randomized algorithm on Max-Cut instances of different sizes $N \in [6, 8, 10, 15, 30, 50, 70, 100, 150]$. Both QAOA and the randomized algorithm return a sample from a certain probability distribution (built by a quantum circuit and uniform distributions respectively). In order to compute the exact optimum $C_{opt}$ we use brute-force search for Max-Cut on small instances (with up to $30$ nodes) and the dynamic program algorithm presented in Ref.\,\cite{Sahni1976} for the initial scheduling problem on bigger instances. \\

\begin{figure}[h!]
\begin{center}
\includegraphics[width=0.7\linewidth]{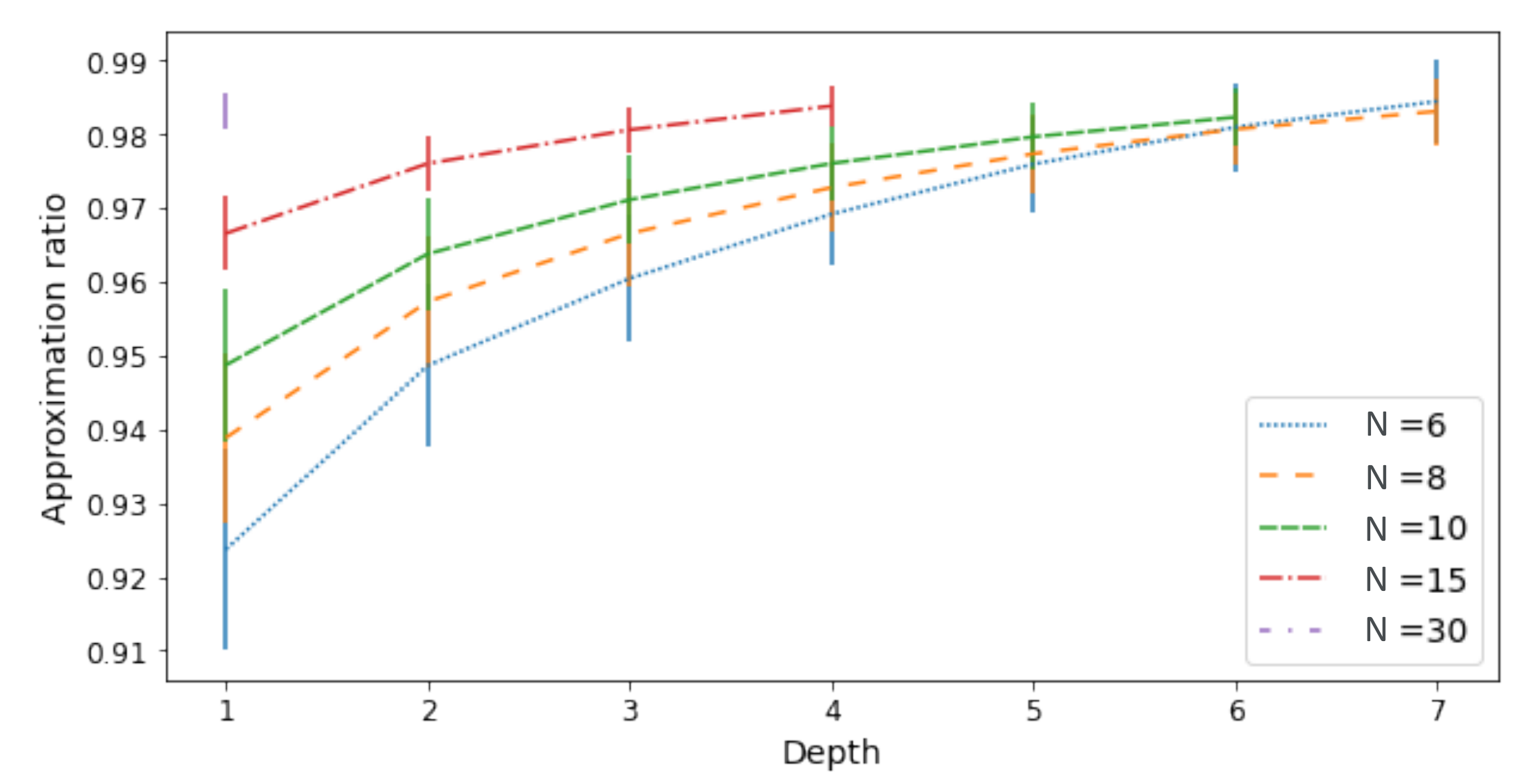}
\includegraphics[width=0.7\linewidth]{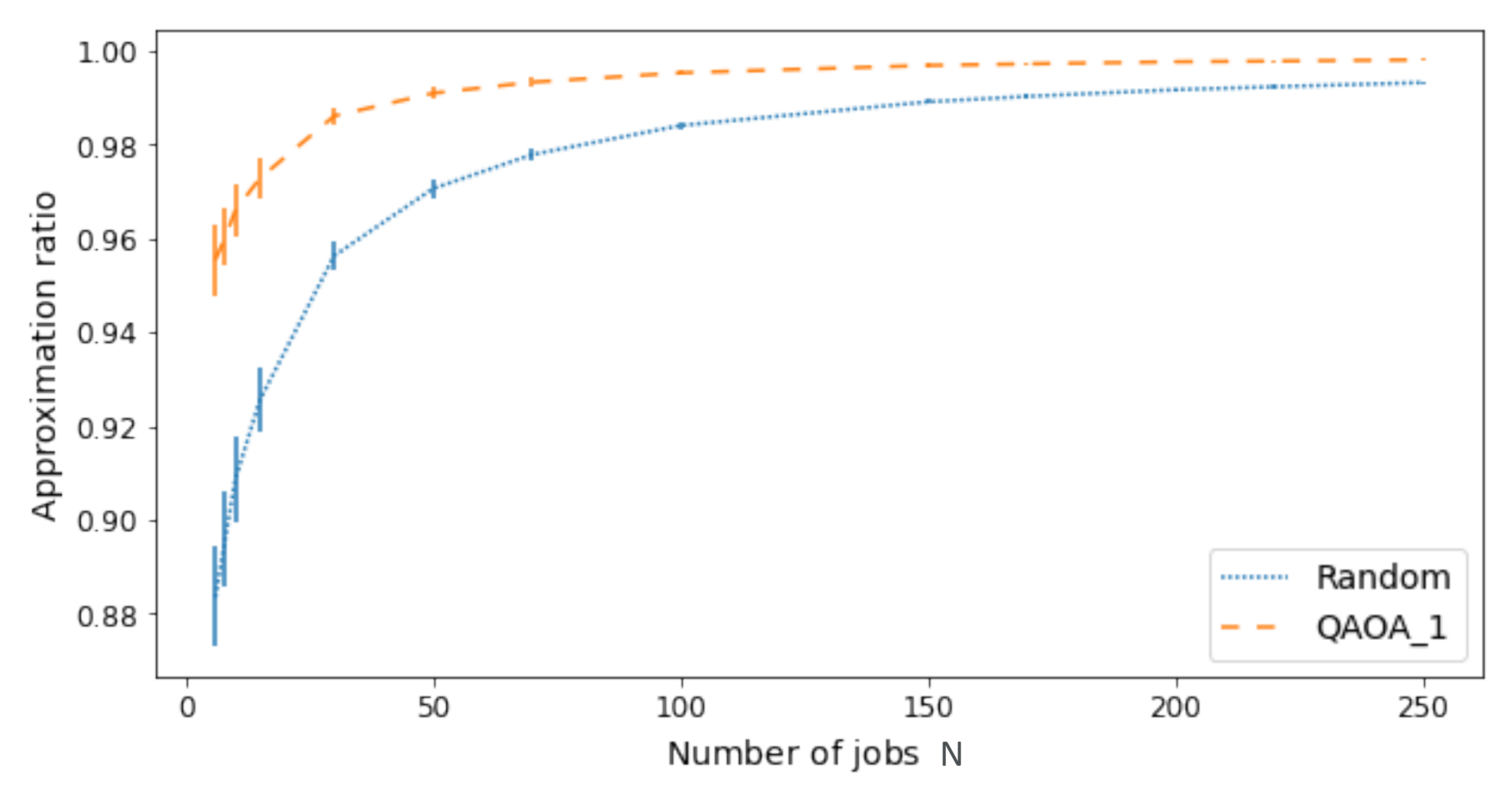}
\end{center}
\caption{Top panel: Evolution of the approximation ratio of $QAOA$ with depth $p$ for the Max-Cut problem. Bottom panel: Evolution of the average approximation ratio with the instance size for $QAOA$ at depth $p=1$  (dashed orange line) and for the randomized algorithm on the \textit{initial scheduling problem} (dotted blue line).}
\label{fig:scaling}
\end{figure}

As expected, we observe on the top panel of Fig. \ref{fig:scaling} that, for a fixed value of $N$, the approximation ratio improves with the QAOA depth $p$. \\

Surprisingly, we also notice that the approximation ratio gets better with the size of the Max-Cut instance. Such behavior is also observed for the randomized algorithm, as illustrated on the bottom panel of Figure \ref{fig:scaling}. In this numerical experiment we observe that QAOA finds better solutions than the randomized algorithm.

\begin{figure}[h!]
\begin{center}
\includegraphics[width=0.7\linewidth]{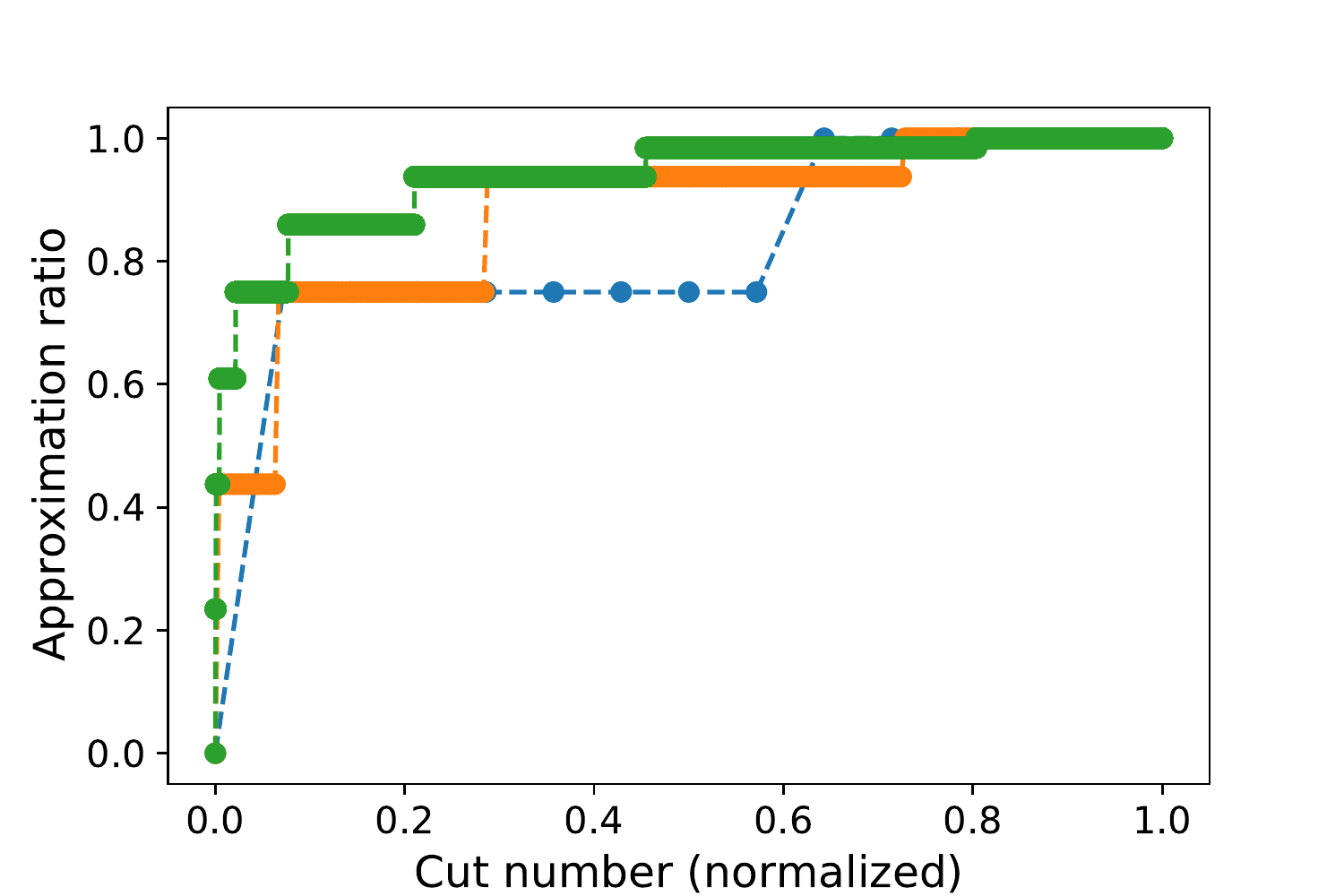}
\end{center}
\caption{Approximation ratios of all possible cuts in the unweighted Max-Cut problem on complete graphs of size $N$, sorted by increasing values. The blue, orange and green dots correspond to $N=4$, $N=8$, and $N=16$, respectively. }
\label{fig:combinatorial}
\end{figure}

The good performances of the randomized algorithm shown on the bottom panel of Figure\,\ref{fig:scaling} suggest that the difficulty of the problem under consideration decreases with its size. To confirm this insight, let us consider the simpler case of unweighted Max-Cut on a complete graph. In this scenario, choosing a cut at random gives an approximation ratio of $\left[ \sum_k \binom{N}{k}k(N-k)\right]/(N^2/4)$ which goes to $1$ in the large $N$ limit. This fact is illustrated in Fig.\,\ref{fig:combinatorial}, where we show the approximation ratios of all possible cuts for different values of $N$. This plot shows that, for a fixed positive value of the normalized cut number, the corresponding approximation ratio approaches 1 in the large $N$ limit. While our findings suggest that the presence of $\mathcal{O}(1)$ weights in our problems leads to the same behavior, it does not extend to instances in which the magnitude of the weights would increase with $N$.   \\

\begin{figure}[h!]
    \centering
    \includegraphics[width=\textwidth]{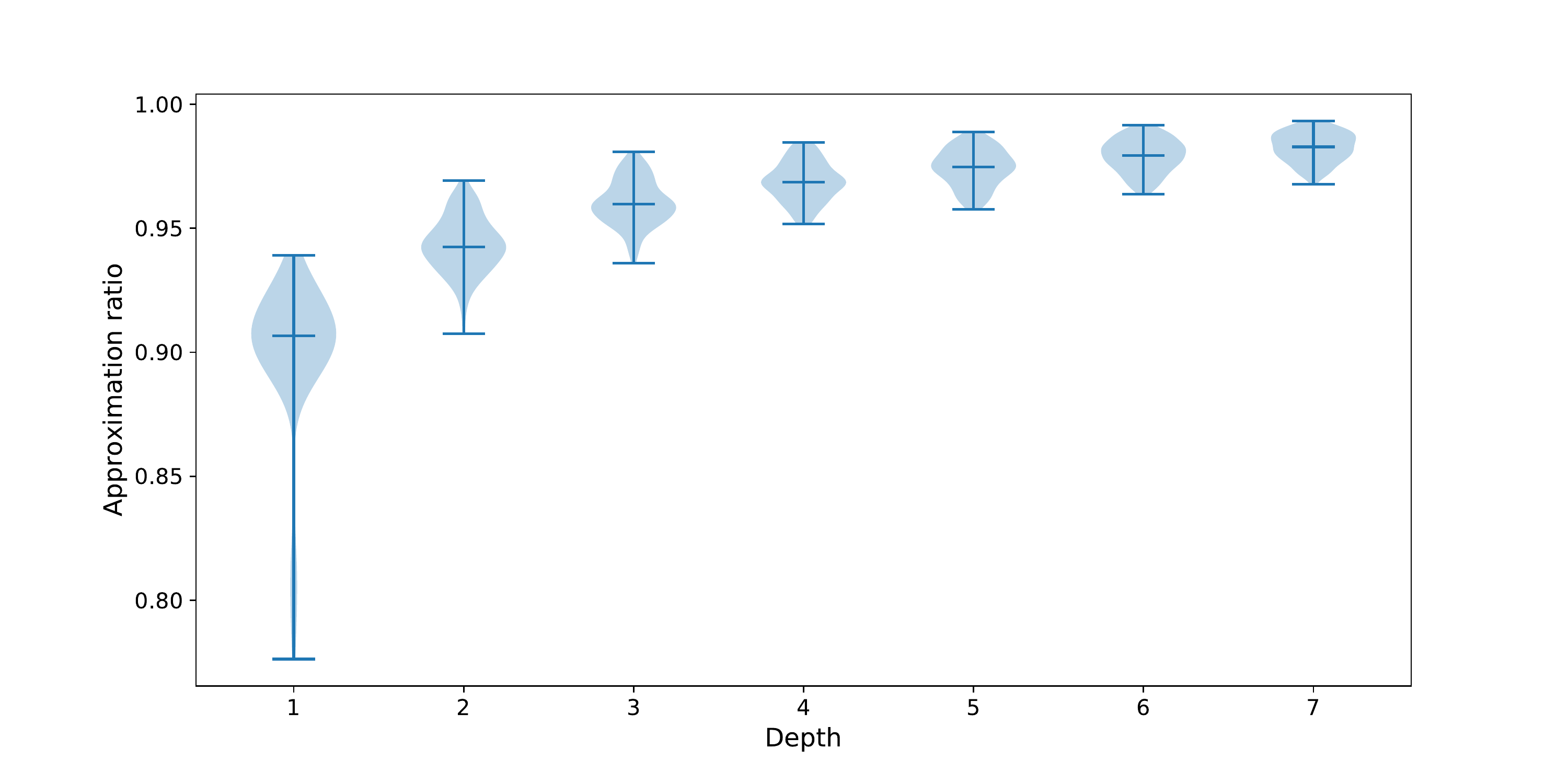}
    \caption{\textbf{Max-4-Cut}. QAOA for $p=7$ layers, and the statistical distribution of the approximation ratio achieved for 98 instances. Proper re-weighting of the graph uniforms the instances. The high value of the approximation ratio achieved is an encouraging result, showing that quantum approaches are comfortably higher than the classical approximation minimal guarantee of 0.857487 \,\cite{de2004approximate}, even at low depths. While initial optimization might not be ideal, as suggested by the tail of distribution on the first layer, it is corrected by the global smoothing at the next layer.}
    \label{fig:M4C}
\end{figure}

We now present numerical results for Max-4-Cut. Using the normalization factor introduced in Appendix \ref{Param_opt}, we ran QAOA for $p=7$ layers, and plotted the statistical distribution of the approximation ratio achieved for 98 different instances (Fig. \ref{fig:M4C}). As expected, the approximation ratio increases to 1 with the number of layers. The high value of the approximation ratio achieved is an encouraging result, showing that quantum approaches are comfortably higher than the classical minimal approximation guarantee of 0.857487 \,\cite{de2004approximate}, even at low depths. At $p=1$, the tail of the distribution indicates that some instances have been poorly optimized. The fact however that this tail disappears in the next layers shows that initial poor optimization can be corrected in the following layers. This is due to the fact that at the end of each layer our classical algorithm implements a rapid local optimization on all parameters.

\subsubsection{Optimal Scheduling of Load Time Intervals within Groups (SC2)}\label{QMISRes}

Previous studies on quantum approaches for solving UD-MIS investigated the influence of quantum noise QAOA\,\cite{Henriet20a}, and compared the performances of Quantum Annealing procedures\,\cite{Serret20} to classical approximation algorithms\footnote{Note that, for the small instances studied here, classical brute-force algorithm can find the exact solution in a short time}. \\

\begin{figure}\centering
\subfloat[Evolution of the approximation ratio for 84 instances of Unit-Disk MIS ($N=15$).]{\label{c}\includegraphics[width=\textwidth]{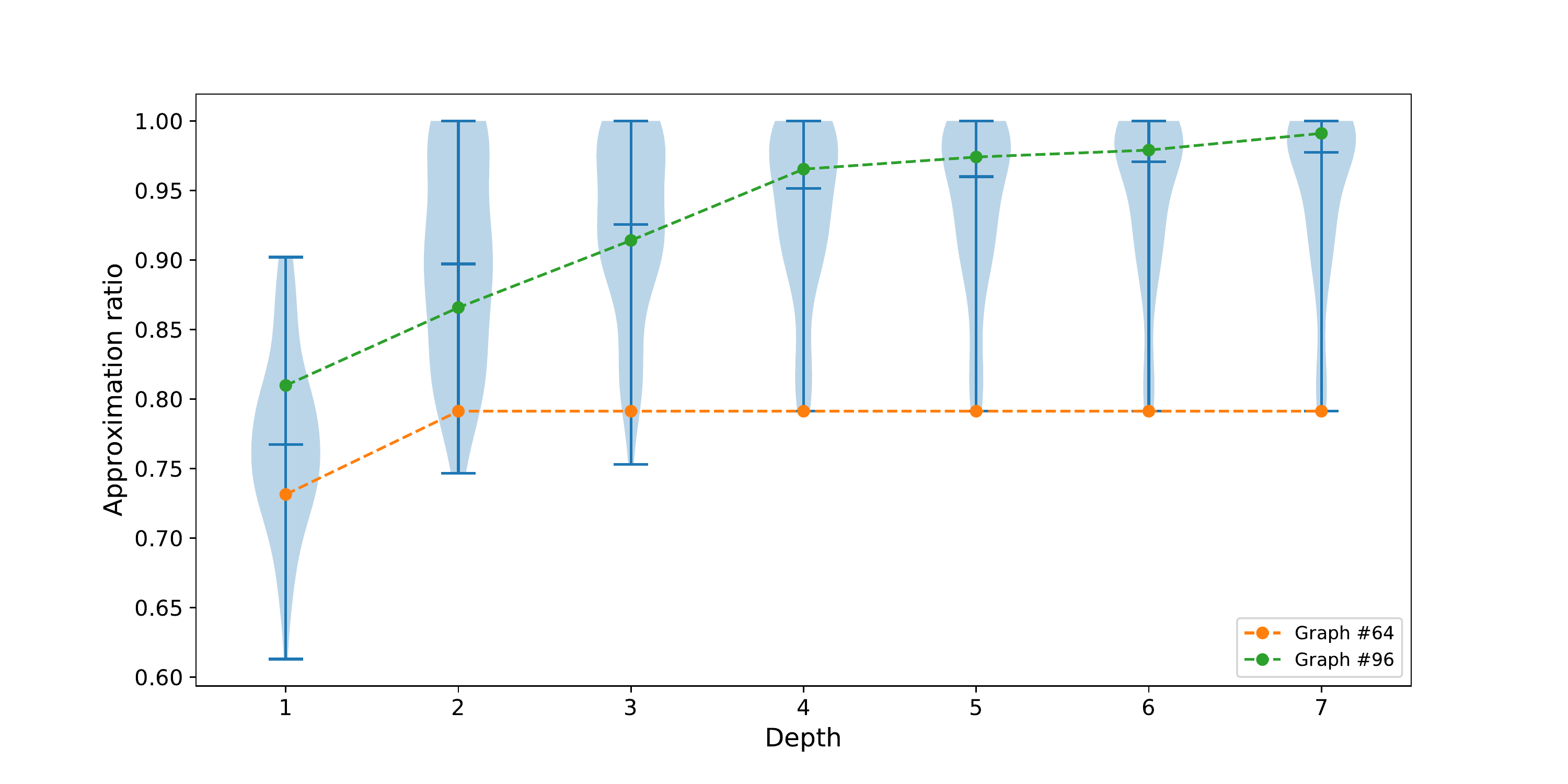}}\newline
\subfloat[Graph $\#$96. ]{\label{a}\includegraphics[width=.3\linewidth]{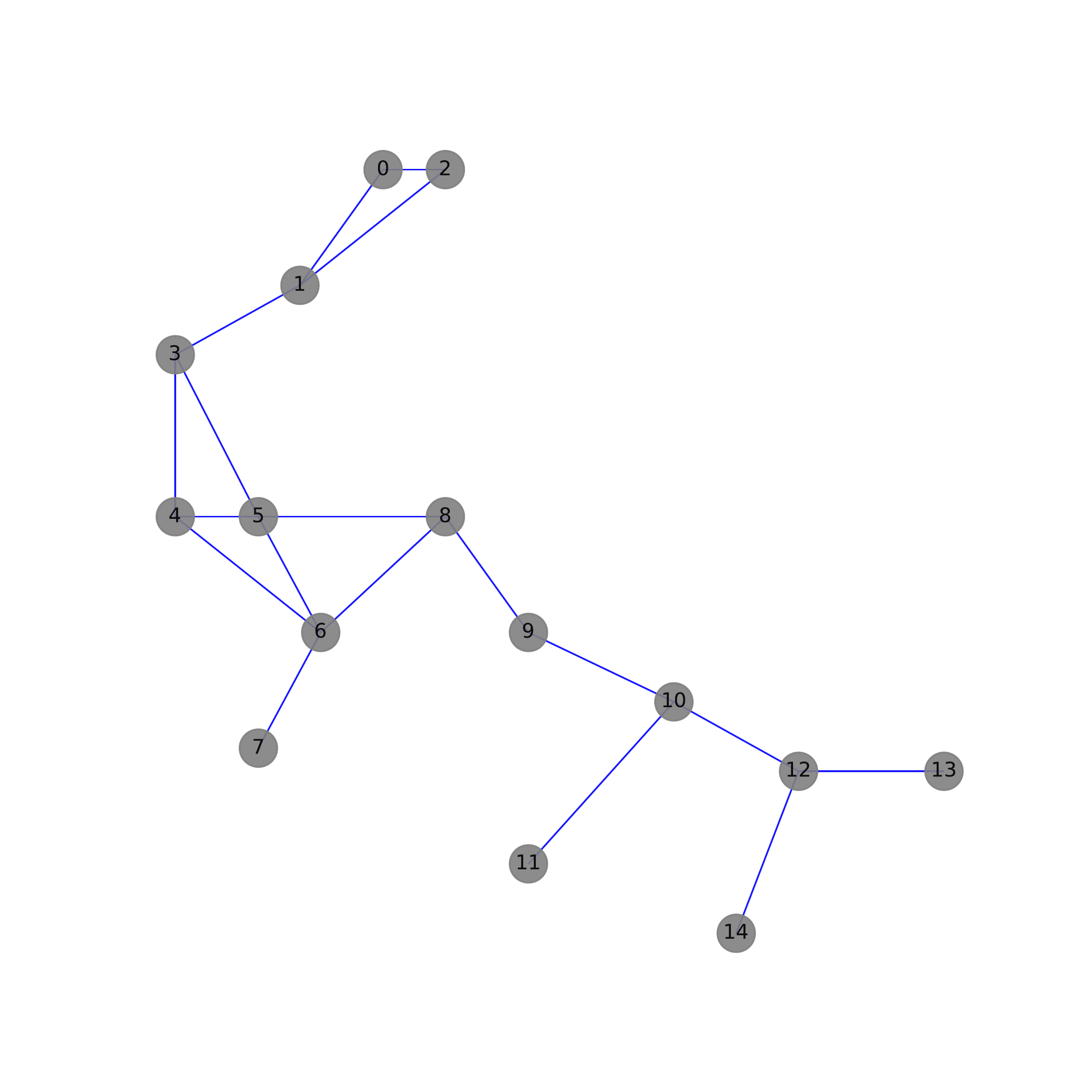}}
\subfloat[Graph $\#$64.]{\label{b}\includegraphics[width=.3\linewidth]{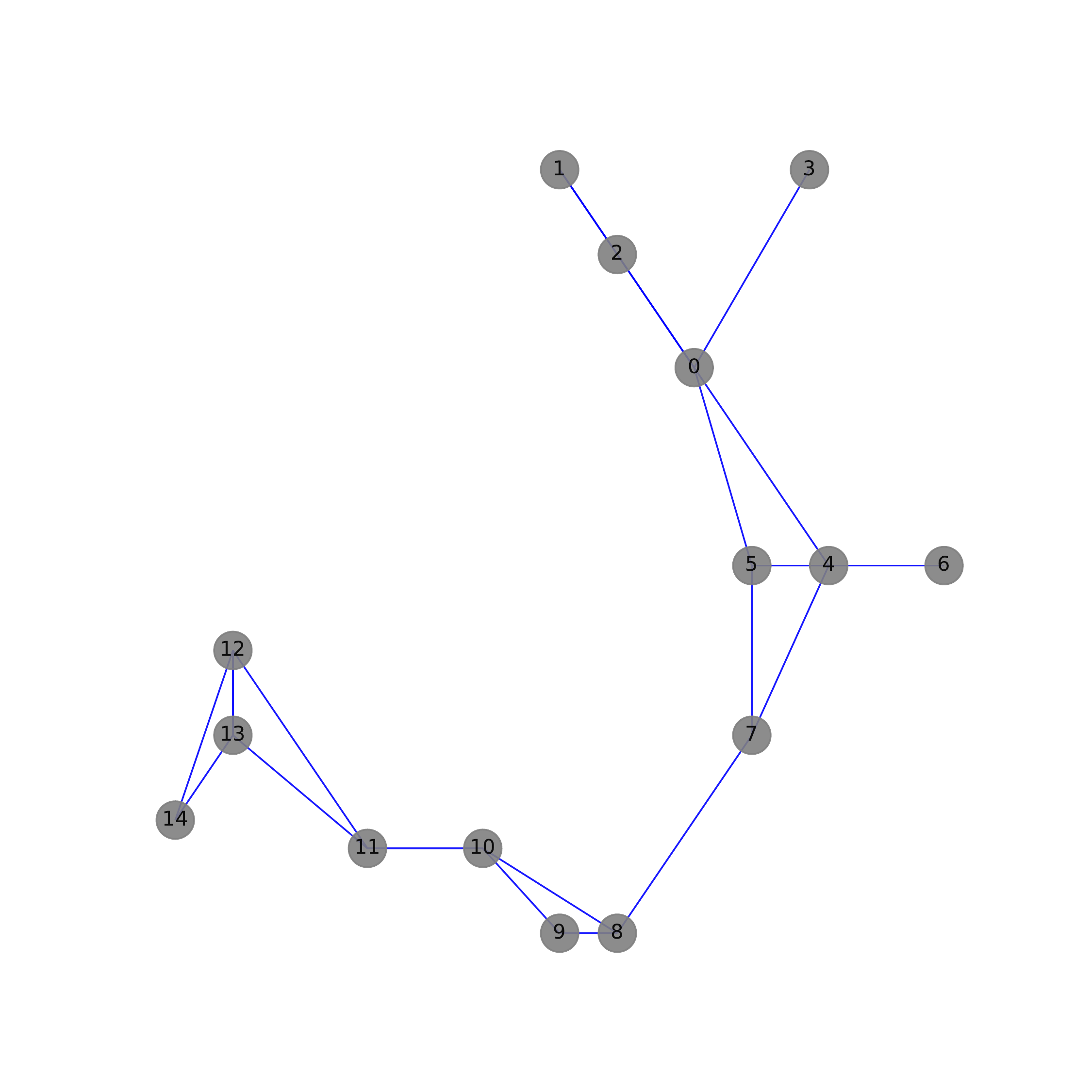}} 
\caption{\textbf{(a)} For each layer, the median is plotted and the violin envelop represents the distribution of the approximation ratio achieved for each of the 84 points. While the approximation ratio grows rapidly to 1 in most cases, there seems to be some instances that are harder to optimize. We isolate a graph representative of the best-case bulk \textbf{(b)} and one for the worst-case bulk (\textbf{c)}. Further understanding of the worst-case graphs  is important in order to appreciate the quality of a quantum approximation algorithm.}
\label{fig:MIS}
\end{figure}

For the classical loop of QAOA, we use here a new global optimization process (Egg optimization) described in the Appendix \ref{appendix:Egg} that uses a global optimization process called differential evolution (DE)\,\cite{Storn1997}. Using (DE) enables to escape from local minima quite easily. Another important advantage is that the function evaluations can be done in parallel. More specifically, we ran our program on the OCCIGEN supercomputer based in Montpellier, France, where we used 28 CPU cores in parallel for our calculations. This reduced by a factor of 8.8 the time of derivation. In addition, we reduced the amount of phase space addressed by making educated guesses from layer $p-1$ to layer $p$, strongly inspired by Ref.\,\cite{zhou2018quantum}, and demonstrated precise results with much fewer function calls. The combination of educated guesses and parallel function evaluations made for a consequent speed-up, bringing down typical calculation times of half a day to an hour. Finally, the global optimizer (DE) has shown strong results in the context of noisy and changing landscape \cite{Rocca2011}, a typical behavior of our noisy intermediate-scale quantum platforms. Using (DE) might prove robust in an experimental setup.\\

As can be seen in Fig.\,\ref{fig:MIS}(a), the performances of QAOA for solving the UD-MIS problem are good in average, exceeding approximation ratios of $0.95$ after seven QAOA layers. As can be noted in the figure, the distribution of the approximation ratio for each layer is rather wide. At the third layer, the distribution starts to separate in two bulks. The lower bulk stagnates by the fourth layer as the approximation ratio stays inferior to 0.85 until the last layer, while the approximation of the upper bulk increases to one as the depth grows. For completeness, we show on panel (b) and (c) of Fig.\,\ref{fig:MIS} a typical graph instance of each group. The instances in the lower bulk correspond to worst-case scenarios and represent $9.5\%$ of the instances. Understanding the characteristics of the worst instances is of crucial importance to characterize the quantum approaches. Indeed in the approximation theory the quality of an algorithm is benchmarked on worst-case instances. The approximation ratio achieved by the algorithm on these particular instances is a guarantee from below for any other instance. Finding worst-case scenario has been investigated in the past for Max-Cut on uniform 3-regular graphs at depths $p=\{1,2,3\}$ \,\cite{Farhi14,wurtz2020bounds}. Obtaining a lower bound guarantee of QAOA on UD-MIS, which we leave for future work, would enable us to compare the quantum approach to the classical approximation scheme to assess an eventual quantum advantage.

\section{Conclusions and Perspectives}\label{future}

Qualifying quantum algorithms on difficult optimization problems is of great importance to evaluate the benefit of quantum computing, as these problems are at the core of many industrial applications where they often constitute performance bottlenecks. \\

Two major principles must be implemented in such a process:
\begin{enumerate}
\item Rely on a collaboration between experts in the application field under study and quantum computing experts, in order to design fine-tuned, \textit{ad-hoc}, software-hardware solutions;
\item Benchmark quantum algorithms not only against exact classical algorithms, by nature exponential on this class of problems, but also versus available approximate polynomial ones.
\end{enumerate}
 
The first point is crucial, because quantum solutions can often take advantage of the specific characteristics of the
targeted quantum hardware. The second point is required for a fair-play competition between quantum and classical approaches for difficult optimization, in order to precisely evaluate a potential quantum advantage. 

This paper reports a case study based on this protocol in the field of smart-charging of electric vehicles. We specified two smart-charging problems,  which, although stylized to be treated by available quantum approaches, stay representative of the real operational problems currently solved by the EDF subsidiaries involved in the field. We developed a hardware-efficient implementation of QAOA on quantum devices based on Rydberg atoms arrays to solve these two problems, respectively modeled as "sub-difficult" instances of Max-k-Cut and MIS NP-hard problems. We have experimented these implementations on a real data-set of 2250 loads, and compared quantum solutions to classical approximate ones, up to current limits of classical simulation of quantum hardware with $N\leq 20$ qubits. In both cases, quantum algorithms behave correctly, obtaining high approximation ratios, coherently with the fact that both applications are modeled as "less difficult" instances than the worst-case ones of these two NP-Complete problems.\\

These results, obtained through a rigorous protocol, are very encouraging. Future works will involve testing the quantum approaches on the real Rydberg atoms quantum processor developed by Pasqal in the 100-1000 qubits range\,\cite{Henriet20}; making the smart-charging problems more realistic by incorporating new constraints (e.g maximal available power on the load station), a real challenge as this should make the associated Hamiltonians to be implemented on the processor more complex; more specifically from an application viewpoint, looking for efficient heuristics to transform general graphs in unit-disk ones, which would drastically simplify the procedure for quantum solving of MIS. On this latter point, another interesting option is to explore smart-charging problems which are "naturally" two dimensional -- e.g.~based on the "autonomy radius" of vehicles or the "action radius" of charging points --, thus replacing the costly and hypothetical resolution of the (UD) problem by a simple scale reduction in the plan.

\section*{Acknowledgments}
We thank Thomas Ayral, Antoine Browaeys, Thierry Lahaye and Christophe Jurczak for discussions. 
This work was developped as a collaboration within the European Commission in the Horizon 2020 FET-Quantum Flagship project PASQuanS (817482). It was granted access to the HPC resources of CINES under the allocation 2019-A0070911024 made by GENCI. It was also supported by EDF R\&D, the Research and Development Division of Electricité de France and by Loria, the Lorraine Research Laboratory in Computer Science and its Applications University of Lorraine, France.

WL was supported by the Austrian Science Fund (FWF) through a START grant under Project No. Y1067-N27 and the SFB BeyondC Project No. F7108-N38, the Hauser-Raspe foundation. This material is based upon work supported by the Defense Advanced Research Projects Agency (DARPA) under Contract No. HR001120C0068. Any opinions, findings and conclusions or recommendations expressed in this material are those of the author(s) and do not necessarily reflect the views of DARPA. 

\newpage
\appendix
\section{Mapping interval graphs to unit-disk graphs of Rydberg atoms on the quantum machine }\label{AnnexQUD}
The (UD) problem introduced section \ref{FromIG2UDG} can be solved by a classical reformulation-linearization of non-convex quadratic programs into binary/integer ones, by considering variables $x_i$ and $y_i$ as integers and replacing them by their binary expansions in constraints(\ref{CtrDist}) and (\ref{CtrDistComp}), then adding  binary variables and appropriate constraints to represent the products between them \cite{Park17,Elloumi19,Billionnet08}.
The resulting binary/integer linear model is as follows:

\begin{flalign}
&Min_{x_q,y_q}\lbrace L \rbrace\\
&s.t.\nonumber\\
&\forall i \in V,x_i=\sum_{k=0}^{k=log(\overline{L})}2^kbx_i^k; \quad y_i=\sum_{k=0}^{k=log(\overline{L})}2^kby_i^k\label{binY}\\
&\forall i \in V, x_i < L; y_i < L\\
& L \leq \overline{L}\\
\\
&\forall \langle i,j \rangle \in E:&&\\
&\rho^2 \le X_jX_j -2 X_iX_j + X_iX_i + Y_jY_j -2 Y_iY_j + Y_iY_i \le r^2\\
& X_iX_j=\sum_{k=0}^{k=log(\overline{L})}\sum_{k'=0}^{k'=log(\overline{L})}2^{(k+k')}wx_{i,j}^{k,k'}\\
& Y_iY_j=\sum_{k=0}^{k=log(\overline{L})}\sum_{k'=0}^{k'=log(\overline{L})}2^{(k+k')}wy_{i,j}^{k,k'}\\
& wx_{i,j}^{k,k'} \le bx_i^k; \quad wx_{i,j}^{k,k'} \le bx_j^{k'}; \quad wx_{i,j}^{k,k'} \ge bx_i^k + bx_j^{k'}-1\\
(UDRLT)\qquad&wy_{i,j}^{k,k'} \le by_i^k; \quad wy_{i,j}^{k,k'} \le by_j^{k'}; \quad wy_{i,j}^{k,k'} \ge by_i^k + by_j^{k'}-1\\
\\
&\forall \langle i,j \rangle \in \overline{E}:\\
& X_jX_j -2 X_iX_j + X_iX_i + Y_jY_j -2 Y_iY_j + Y_iY_i > r^2\\
& X_iX_j=\sum_{k=0}^{k=log(\overline{L})}\sum_{k'=0}^{k'=log(\overline{L})}2^{(k+k')}wx_{i,j}^{k,k'}\\
& Y_iY_j=\sum_{k=0}^{k=log(\overline{L})}\sum_{k'=0}^{k'=log(\overline{L})}2^{(k+k')}wy_{i,j}^{k,k'}\\
& wx_{i,j}^{k,k'} \le bx_i^k; \quad wx_{i,j}^{k,k'} \le bx_j^{k'}; \quad wx_{i,j}^{k,k'} \ge bx_i^k + bx_j^{k'}-1\label{RLTC1}\\
& wy_{i,j}^{k,k'} \le by_i^k; \quad wy_{i,j}^{k,k'} \le by_j^{k'}; \quad wy_{i,j}^{k,k'} \ge by_i^k + by_j^{k'}-1\label{RLTC2}\\
\nonumber\\
& bx_i^k, by_i^k, wx_{i,j}^{k,k'}, wy_{i,j}^{k,k'} \text{all binary}.
\end{flalign}

Once rewritten this way, and assuming that it is feasible, (UD) can be solved to optimality by conventional Branch and Bound/Cut algorithms, but with an exponential computing time in the number of integer/binary variables, in the worst case. This renders this formulation too costly even for small sized graphs, as for each couple $(x_i,y_i)$ of continuous variables modeling a vertex in the original problem we introduce $2log(\overline{L})$ binary variables $(bx_i,by_i)$ to expand its integer formulation, and $(log(\overline{L}))^2$ binary variables $wx_{i,j}^{k,k'}$/$wy_{i,j}^{k,k'}$ to express each product in the constraints related to the edges.\\

As an alternative, we used a more compact linear formulation which, although not providing guaranty to find an existing solution in the elapsed time allowed, did perform well on the majority of smart-charging graphs tested. Basically, the idea is to linearize the problem by replacing the quadratic constraints by linear ones, expressing the belonging to the rings defined by $(r,\rho)$ by means of a given set of parametrized radius. The resulting  model is as follows:

\begin{flalign}
&Min_{x_q,y_q}\lbrace L \rbrace\\ 
&s.t.\nonumber\\
&\forall i \in V, x_i < L; y_i < L\\
& L \leq \overline{L}\\
(UDR_\Phi)\qquad&\forall \langle i,j \rangle \in E: \bigvee_{\phi \in \Phi}f_\phi(x_i,y_i,x_j,y_j)\label{OR}\\.
&\forall \langle i,j \rangle \in \overline{E}: (|x_j-x_i| > r) \vee (|y_j-y_i| > r)\label{RING}
\end{flalign}

where $f_\phi$ is a set of clauses enforcing the belonging of the point $(x_j,y_j)$ to the radius defined by the angle $\phi$ intersecting the ring of center $(x_i,y_i)$ defined by $(r,\rho)$. For example:
\[
f_{\phi \in [0,\frac{\pi}{2}[}=(\rho\cos\phi \leq x_j-x_i) \wedge  (x_j-x_i \leq r\cos\phi) \wedge (\rho\sin\phi \leq y_j-y_i) \wedge (y_j-y_i \leq rsin\phi)
\]
Because of the "or" and "absolute value" terms in the formulation, this model is still combinatorial, but much more efficient than the previous one.

It is worth noting that in both formulations, it is possible to relax the constraint expressed in the objective function enforcing to locate the Rydberg atoms in the  smallest square contained in $\overline{L} \times \overline{L}$; one thus  obtains a pure "constraint programming" problem, whose solving can stop at the first solution encountered.\\

Both formulations were implemented and tested with Ibm Cplex solver. The second one provided in one hour of elapsed time with 57 instances to be implemented for UD-MIS search on the Rydberg atom based quantum machine, starting from 100 instances of real smart-charging graphs of 15 nodes each.
\section{Optimizing the classical loop of QAOA}
\label{Param_opt}

Although translation of a combinatorial problem focuses on the quantum loop of the QAOA, the quality of the classical optimizer part is of crucial importance. Indeed, an inefficient classical optimizer will bring excessive overhead to the whole process. For practical use, it is of crucial importance to use an optimizer tailored to the machine. In the following paragraphs, we present results on the research for the best classical optimization process. \\

\begin{figure}
    \centering
    \includegraphics[width=\linewidth]{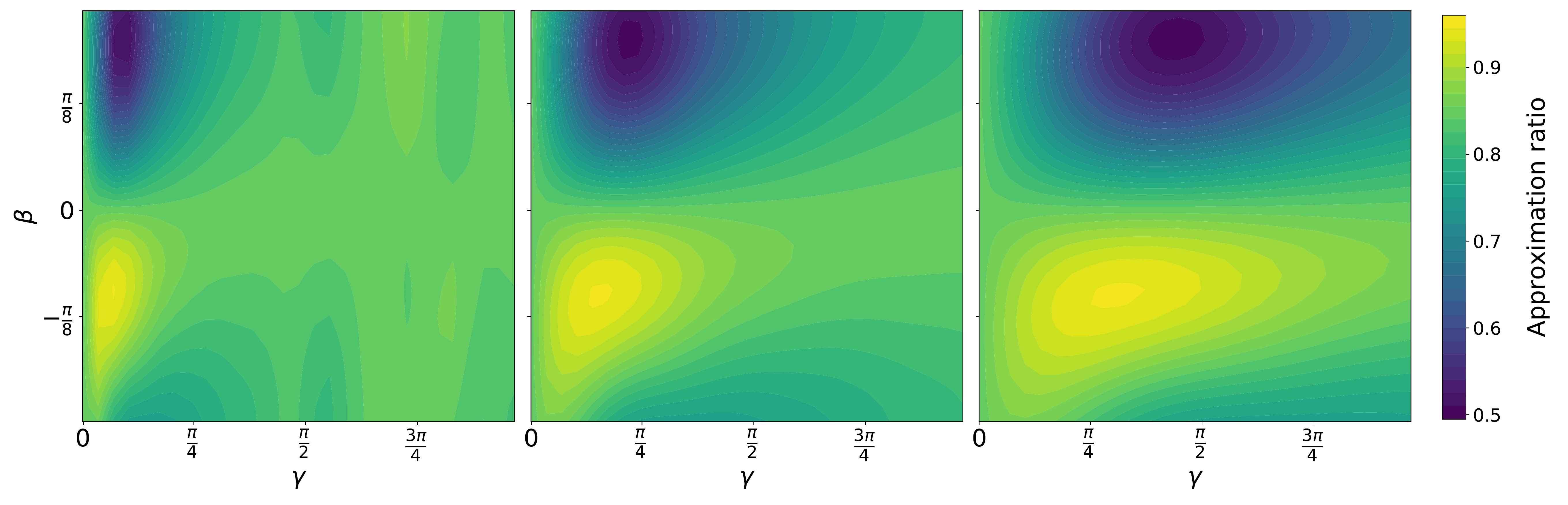}
    \caption{\textbf{The zoom effect}. From left to right, we normalized an instance from (SC1) respectively by $R/5$, $R/2$ and $R$ to observe the effect on the energy landscape of $QAOA_1$.The closer to $R$, the better the zoom on the global minima. The re-weighting of a graph affects the energy landscape: it can therefore be used as leverage to either zoom on the point of interest to apply local optimization, or on the contrary it might be used to zoom out of barren plateaus to explore more interesting phase spaces.}
    \label{fig:normalize_weights}
\end{figure}

In the (SC1) problem, the weights $w_{uv}$ of the edges in the graph are imposed by real data. These weights appear in eq.(\ref{eq:analytical_maxcut_depth_1}) and multiply the parameterized angle $\gamma$. They therefore impact the phase that is applied to each basis state of the computational basis in the QAOA cycle. To visualize the impact of weights, we plot the energy (or cost-function) landscape at level $p=1$ as a function of the parameters $(\gamma,\beta)$. Fig.\ref{fig:normalize_weights} contains numerical simulation of the energy landscape for the Max-Cut problem, evaluated on a graph of size 10, with a resolution of 30 points along each axis. For the same problem, we re-weight the adjacency matrix of the graph by different factors. On the first figure, a high density of peaked valleys and hills indicates that optimization is difficult and would require an important amount of function evaluations to find a decent solution. The amount of peaks and valleys is due to the fact that the cost Hamiltonian adds a phase term to each basis vector $z_i$. Indeed, applying $\hat{C}$ with an angle $\gamma$ modifies $|z_i\rangle$ to $ e^{i\gamma C_i}|z_i \rangle$, where $C_i$ corresponds to the cost of the coloring $z_i$. Modifying the graph weights consequently modifies the the phase applied to each basis state $z_i$. A smaller $C_i$, as seen in the third figure, smooths the cost-function landscape enabling adequate local optimization.  Artificially reducing $C_i$ too much however might over-smooth the landscape, reducing the possible phases that QAOA can apply, hence missing the global minima. We want $C_i$ to be big enough to allow basis states $z_i$ to acquire a phase $e^{i\phi}$ in a comfortable range. At the same time, we do not believe a phase $ \phi > 2 \pi$ is necessary. The re-weighting $C_i = R^* C_i$, where $R^* =\frac{2\pi}{\max(C_i)}$ satisfies the two previous conditions.  Numerically, we find that re-weighting the adjacency matrices of all instances by the factor $R$ indeed concentrates all optimal parameters in a restricted zone of parameter space.

Practically of course, calculating $R^*$ implies knowledge of $\max{C_i}$, which corresponds to the exact solution to the initial problem. We therefore propose the upper bound $R =w_{\max} \frac{N^2}{4}$, where $w_{\max} = \max_{u,v}w_{uv}$. It is calculated from the best-cut on a complete graph where all weights would be equal to $w_{\max}$.

In the scenario of (SC1), we see that the choice for the optimizer depends on the normalization of the weights. If one manages a good normalization, then local methods of optimization can be used with high guarantees of success. In general, we advocate the use of a global optimizer if there is no prior knowledge of the energy landscape. \\

\section{Finding correct parameters: the Egg optimization and local vs. global methods}
\label{appendix:optimization_methods}

\subsection{Educated global guess (Egg) optimization}
\label{appendix:Egg}
In order to find the best variational parameters for $p$ layers, we develop a method based on the idea of making an educated guess from previous layers to the new one  \,\cite{zhou2018quantum}. The educated global guess (Egg) optimization process uses the differential evolution (DE) \cite{Storn1997} rather than a local optimization in an attempt to find the global optima in a wrinkled energy landscape. (DE) works by starting with an ensemble of points in the phase-space, called the \textit{agent population}. Then, theses agents are moved around by recombining their coordinates, and the function is evaluated for these new agents. If the new position brings an improvement, it is kept, otherwise it is discarded. This process is repeated until convergence to a minima, although there is no guarantee that the global minima will be found. While it cannot be sure that our method will always work perfectly, the constant growth of the approximation ratio as $p$ increases in our results is a reassuring indicator, as illustrated on the top panel of Fig. \ref{fig:MIS}.
A second major hurdle in the optimization process is the high-dimensionality of the phase space: for $p$ layers, we need to find the global minimum of a space of size $2p$. We strongly reduced the complexity of the problem by making a global educated guess for the optimal parameters at layer $p$ using the optimal parameters found at layer $p-1$ (Alg.\ref{alg:Egg}). The heuristic of an educated guess from the previous layer to the new one was developed in Ref.\,\cite{zhou2018quantum}. Our version uses a global optimizer rather than a local one in an attempt to find the global optima in a wrinkled energy landscape. This very much improved computation time by reducing the amount of phase space addressed, and demonstrated precise results with much fewer function calls than local methods (see Appendix \ref{appendix:optimization_methods}).
The algorithm is described in pseud-code below (Alg.\ref{alg:Egg}). It works as follows: find optimal parameters $(\gamma_1^*,\beta_1^*)$ for $p=1$ . For the next layer, optimize the function $C: (\gamma_2,\beta_2) \mapsto C (\gamma_1^*,\gamma_2,\beta_1^*,\beta_2)$. As such, two variables are already fixed  and the space to explore is once again only bi-dimensional. Once the optimization ends on the two new coordinates, a local optimization is done on all the coordinates. This quick step enables to re-calibrate the previous parameters: it is therefore possible to achieve a trotterization process for high values of $p$.

\begin{algorithm}
    \SetAlgoLined
    \DontPrintSemicolon 
    \SetKwInput{Input}{Input}
    \SetKwInput{Output}{Output}
    \SetKwInOut{KwIn}{Input}
    \SetKwInOut{KwOut}{Output}
    \KwIn{$|z_0 \rangle,\hat{C}, \hat{M}, p$.}
    \KwOut{Optimal parameters $(\gamma_1,\dots,\gamma_p,\beta_1,\dots,\beta_p)$.}
    \hrulealg
    $\gamma_1, \beta_1 \leftarrow  \text{DE}(|z_0 \rangle,\hat{C}, \hat{M})$ \;
    \For {$k$ in range$(2,p)$}{
     $ |z_{k-1} \rangle = e^{i\beta_{k-1}\hat{M}}e^{i\gamma_{k-1}\hat{C}} \dots  e^{i\beta_{1}\hat{M}}e^{i\gamma_{1}\hat{C}} |z_0 \rangle$ \;
     $\gamma_k, \beta_k = \text{DE}(|z_{k-1} \rangle,\hat{C}, \hat{M})$ \;
     \textbf{local optimization}: BFGS($\gamma_1, \dots, \gamma_k, \beta_1, \dots, \beta_k, |z_0 \rangle, \hat{C},\hat{M}$)
     }
    \textbf{return} $(\gamma_1,\dots,\gamma_p,\beta_1,\dots,\beta_p)$
    \caption{\textsc{Educated global guess  (Egg) optimization}}
    \label{alg:Egg}
\end{algorithm}

\subsection{Local vs. global methods}
In Ref.\,\cite{Farhi00} the grid search was proposed to find optimal parameters for QAOA. This method returns a global optima but quickly becomes intractable while growing the depth $p$. Thus a practical approach is to use numerical optimization methods. They may be divided in two groups: global and local methods. Global optimization procedures (such as grid search) address a higher amount of the search space while the local ones return a solution that lays in a neighborhood of a predefined initial point.

An experimental protocol for QAOA should specify the method to use in the parameter optimization step as well as its additional parameters. Local optimization routines are often used in works on applied QAOA\,\cite{Egger2020,Shaydulin_2019,,Alexeev2019}, however the reasons why a certain method is chosen for a particular application are usually omitted. For (SC1) we compared different methods in terms of their performances (evaluated by the approximation ratio of the QAOA with returned parameters) and costs (measured in number of calls of the function to optimize, i.e.~a function that computes $\langle \bm{z}_{\bm{\gamma, \beta}} |\hat{C}| \bm{z}_{\bm{\gamma, \beta}} \rangle$).\\

In order to apply a local optimization routine one has to specify a good initialization point. Without any prior knowledge about the parameter space it is chosen randomly from a set of possible values (usually several points are studied in parallel). However in some applications a pattern in optimal parameters is observed. Such pattern may rely optimal parameter values at different depths, as in\,\cite{zhou2018quantum}, or the values for certain groups of instances\,\cite{brandao2018fixed}. 

\begin{figure}[h!]
\begin{center}

	\subfloat[$QAOA_1$ optimal parameters for instances of different sizes]
	{
	\includegraphics[width=0.5\linewidth]{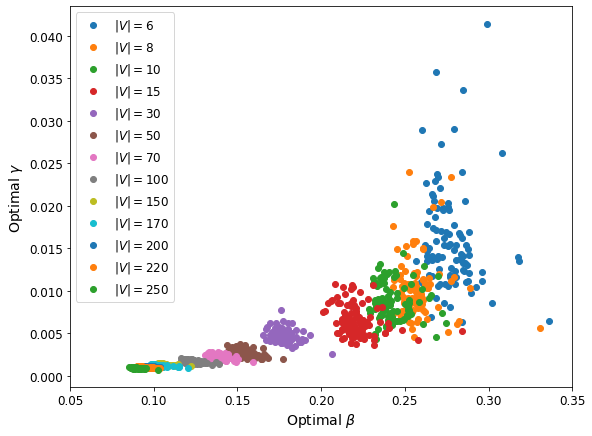}
	}
	\\
	\subfloat[An example of optimal parameters for different $p$ for an instance with $|V|=8$]
	{
	\includegraphics[width=0.9\linewidth]{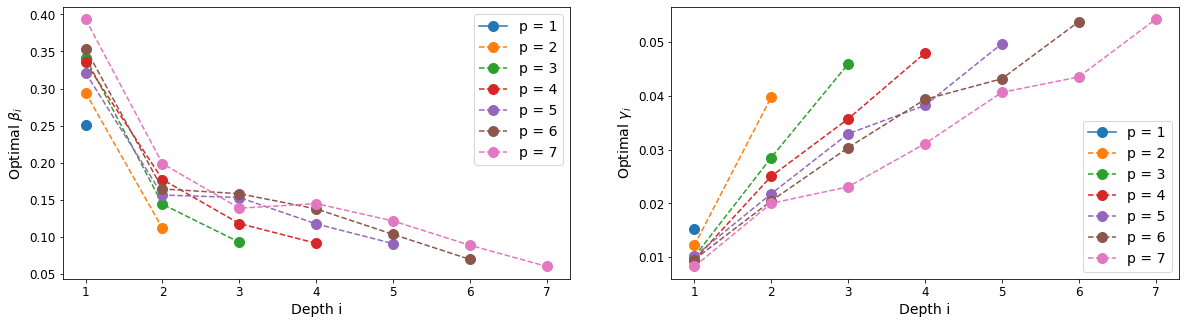}
	}	
\end{center}
\caption{Initialization strategy for numerical optimization: a pattern is observed for $p=1$ (a) and the a linear transformation of parameters at $p-1$ for $p>1$ (b)}
\label{fig:init}
\end{figure}

For $QAOA_1$ the equality $\gamma=0$ implies the absence of any quantum effect (the resulting distribution stays uniform). In Fig. \ref{fig:init} we observe that the absolute values of $\beta$ and $\gamma$ get smaller for instances with bigger $|V|$, but in general it does not imply that the quantum effect becomes less important as $\gamma$ may changed to an arbitrary positive value by multiplying all weights with some factor. We also can't claim that $QAOA_1$ output distribution gets closer to uniform even if we observed that their mean values effectively get closer.\\

We found that at depth $p=1$ optimal parameter values are close for instances with the same number of nodes $|V|$ and that they get concentrated while growing the size of the graph \ref{fig:init}. For $p=1$ we developed an initialization procedure for local methods that relies on the observed pattern. At depths $p>1$ we used the \textit{INTERP} heuristic introduces in Ref.\,\cite{zhou2018quantum} which is based on the intuition that optimal parameters at depth $p$ are close to a linear transformation of optimal values at $p-1$, which seems to be true for our instances (see Fig. \ref{fig:init}).

\begin{figure}[h!]
\centering
\includegraphics[width=0.9\linewidth]{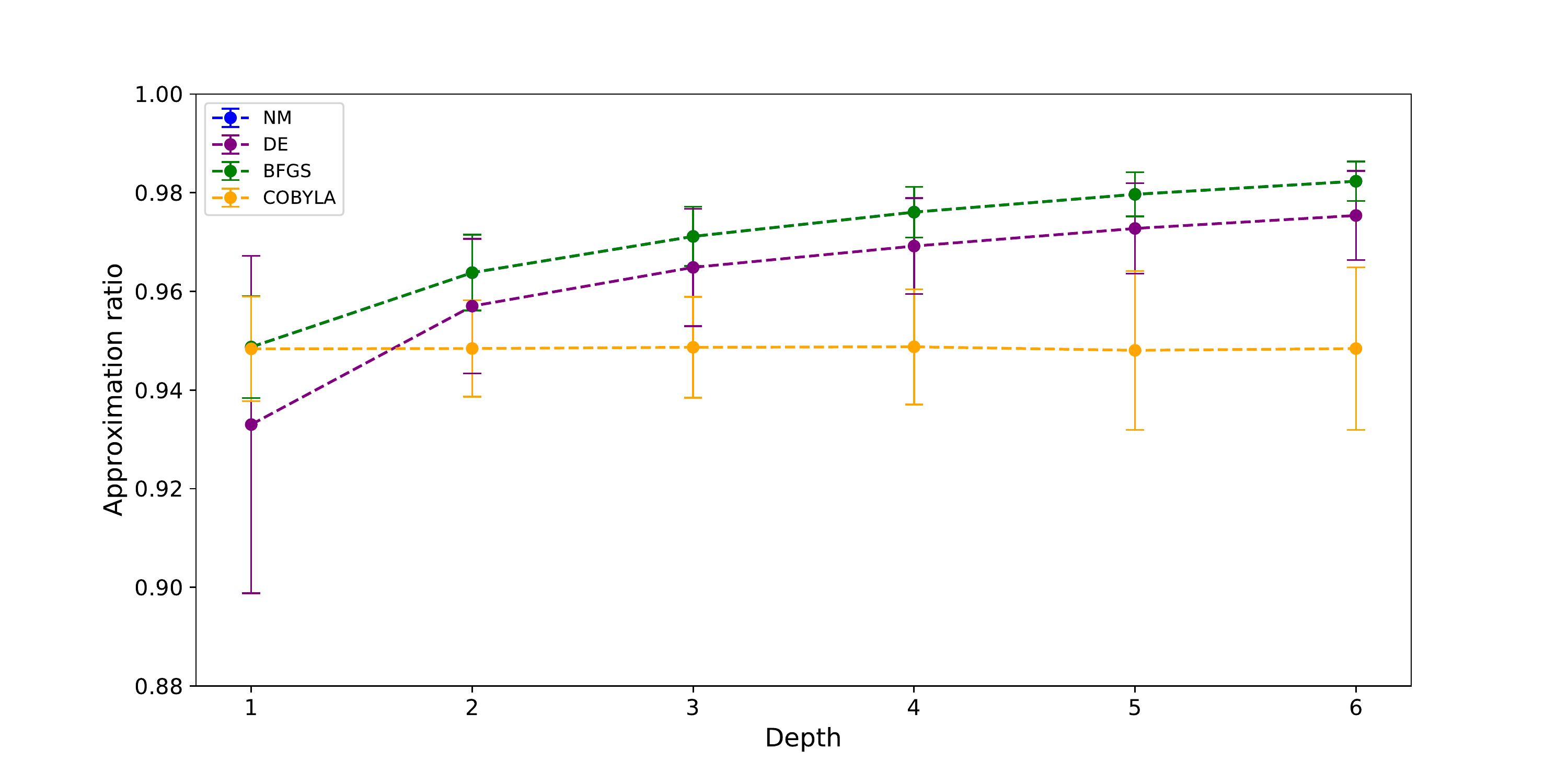}
\includegraphics[width=0.9\linewidth]{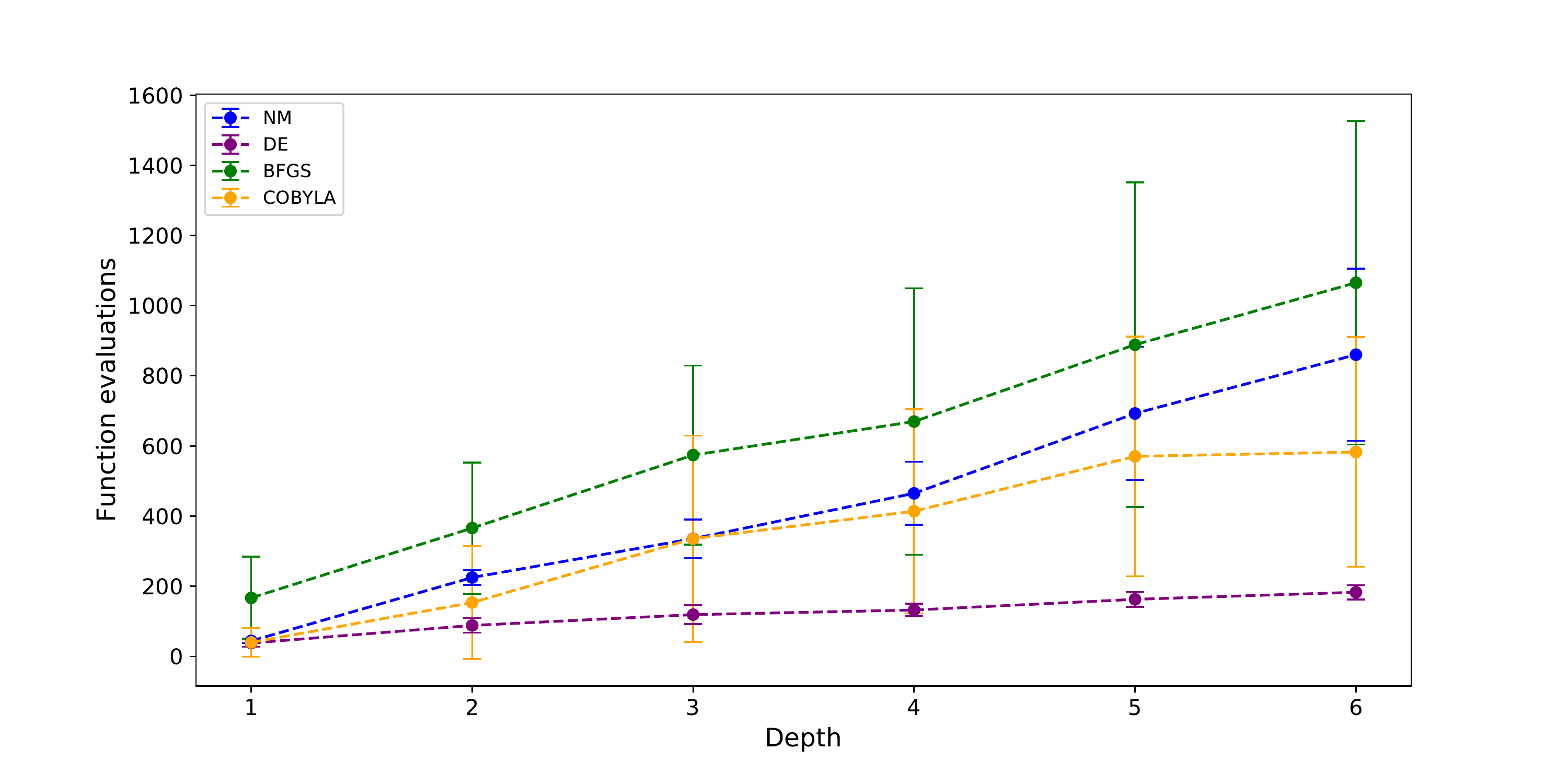}
\caption{Cost and performance of different numerical optimization methods on Max-Cut (SC1), for graphs with size $N= 10$. Differential Evolution (DE) corresponds to a global optimization algorithm, inspired by the INTERP method. While yielding results very close to Nelder-Mead, it does so by requiring very little number of function evaluations. Building optimization processes such as DE that require little function evaluations is key in the NISQ era as it ensures quicker performances on unstable devices.}
\label{fig:comp_methods}
\end{figure}

A choice of the optimization routine has a high impact on the performance and cost of an experimental implementation of QAOA as was shown in Ref.\,\cite{Guerreschi2017}. By experimentally comparing different optimization methods we observed that for local methods, a gradient-free \textit{Nelder-Mead} \cite{Nelder1965} is the best choice for our purposes: its performance is comparable to the one of the quasi-Newton \textit{BFGS} \cite{Gill2019} (and both methods outperform the \textit{COBYLA} routine \cite{Powell1994} which is often used in works on applied QAOA \cite{Egger2020}) while it requires less evaluations of the objective function (see Fig. \ref{fig:comp_methods}).

We compare these local methods with \textit{Differential Evolution} (DE) \,\cite{Storn1997}. We limited the number of function calls while using DE in order to have the best trade-off between global exploration and low number of function evaluation. As seen in Fig.\ref{fig:comp_methods}it is a very effective method, for which the approximation ratio grows close to Nelder-Mead using however much less function evaluations with the layers. It should be kept in mind nonetheless that global optimization methods require tuning hyper-parameters, a process that must be adjusted by hand. The optimal hyper-parameters might change from one problem to another, a reason why they are not so popular. But building optimization processes such as DE that require little function evaluations is key in the NISQ era as it ensures quicker performances on unstable devices.

\printbibliography

\end{document}